\documentclass[11pt]{article}
\usepackage{setspace}
\usepackage{amsmath, hyperref, wasysym}
\usepackage{latexsym}
\usepackage{amssymb}
\usepackage{cite}
\usepackage{amsthm}
\usepackage{lineno}
\usepackage[margin=1in]{geometry}
\usepackage{wasysym}

\usepackage{tabularx}
\usepackage{caption}
\newcolumntype{C}{>{\centering\arraybackslash}X}
\usepackage{lscape}
\usepackage{color,soul}
\usepackage{cancel}
\usepackage{booktabs}
\usepackage{enumerate}
\usepackage{placeins}
\usepackage{enumitem}
\usepackage{adjustbox}

\usepackage{fancyhdr}
\usepackage{graphicx}
\usepackage{times}
\usepackage{sidecap}
\usepackage{color}
\usepackage{multicol}
\usepackage{placeins}
\usepackage{authblk}
\usepackage{arydshln}
\usepackage{float}
\usepackage{dsfont}
\usepackage{caption}
\usepackage{subcaption}
\usepackage{booktabs}
\usepackage{enumerate}
\usepackage{enumitem}
\usepackage{adjustbox}
\restylefloat{table}
\usepackage{lipsum}
\usepackage[labelfont=bf]{caption}
\sidecaptionvpos{figure}{c}

\begin{document}

\title{Forecasting elections using compartmental models of infection}

\author[a]{Alexandria Volkening}
\author[b]{Daniel F. Linder} 
\author[c]{Mason A. Porter}
\author[d]{Grzegorz A. Rempala}

\affil[a]{NSF--Simons Center for Quantitative Biology, Northwestern University, Evanston, IL 60208 (\href{mailto:alexandria.volkening@northwestern.edu}{alexandria.volkening@northwestern.edu})} 
\affil[b]{Medical College of Georgia, Division of Biostatistics and Data Science, Augusta University, Augusta, Georgia 30912 (\href{mailto:dlinder@augusta.edu}{dlinder@augusta.edu})}
\affil[c]{Department of Mathematics, University of California, Los Angeles, Los Angeles, California 90095 (\href{mailto:mason@math.ucla.edu}{mason@math.ucla.edu})}
\affil[d]{Division of Biostatistics, College of Public Health, The Ohio State University, Columbus, OH 43210 (\href{mailto:rempala.3@osu.edu}{rempala.3@osu.edu})}
\maketitle

\begin{abstract}
Forecasting elections --- a challenging, high-stakes problem --- is the subject of much uncertainty, subjectivity, and media scrutiny. To shed light on this process, we develop a method for forecasting elections from the perspective of dynamical systems. Our model borrows ideas from epidemiology, and we use polling data from United States elections to determine its parameters. Surprisingly, our general model performs as well as popular forecasters for the 2012 and 2016 U.S. races for president, senators, and governors. Although contagion and voting dynamics differ, our work suggests a valuable approach to elucidate how elections are related across states. It also illustrates the effect of accounting for uncertainty in different ways, provides an example of data-driven forecasting using dynamical systems, and suggests avenues for future research on political elections. We conclude with our forecasts for the senatorial and gubernatorial races on 6~November 2018, which we posted on 5 November 2018.
\end{abstract}

% REQUIRED

%%%%%

%%%%%%

\section{Introduction}
Despite what was largely viewed as an unexpected outcome in the 2016 United States presidential election, recent work \cite{jennings2018} suggests that national polling data is not becoming less accurate. Election forecasting is a complicated, multi-step process that often comes across as a black box. It involves polling members of the public, identifying likely voters, adjusting poll results to incorporate demographics, and accounting for other data (such as historical trends). The result is a high-stakes, high-interest problem that is rife with uncertainty, incomplete information, and subjective choices \cite{Lauderdale}. In this paper, we develop a new forecasting method that is based on dynamical systems and compartmental modeling, and we use it to help examine U.S. election forecasting.
 
People typically use two primary types of data to forecast elections: polls and ``fundamental data''. {Fundamental data {consists of different factors on which voters may base their decisions \cite{Gelman}}; it includes economic data, party membership, and various qualitative measurements (e.g., how well candidates speak) \cite{SilverBook,HUMMEL2014123}.}  Mainstream forecasting sources, such as newsletters and major media websites, offer varying levels of detail about their techniques and often rely on a combination of polls and fundamental data. Some analysts forecast vote margins at the state or national level (e.g., \cite{538govSen,HuffPostMethod,LATimes,NYTsenate}), while others (e.g., \cite{Sabato,270toWin,cookGov,insideSenGov}) call outcomes by party without giving margins {of victory}. {We will refer to the former approaches as ``quantitative" and the latter as ``qualitative", though both settings typically employ quantitative data.} Among quantitative forecasters, it is important to distinguish between those who aggregate publicly available polls from a range of sources and those who gather their own in-house polls (e.g., \emph{The Los Angeles Times} \cite{LATimes}). For example, FiveThirtyEight \cite{538govSen} is a poll aggregator that is known for its pollster ratings; they weight polls more heavily from sources that they judge to be more accurate \cite{538methodPres}. After adjusting polls to account for factors such as recency, poll sample, {convention bounce\footnote{{Candidates often receive a brief increase (i.e., a ``bounce'') in support after their party's convention. Because of this, some analysts adjust polling data shortly after the Republican and Democrat conventions take place} \cite{538methodPres}.}}, and polling source, FiveThirtyEight uses state demographics to correlate random outcomes, such that similar states are more likely to behave similarly \cite{538methodPres}. {For instance, in one of FiveThirtyEight's simulations, they may adjust the projected vote among Mormons by a few percentage points in favor of the Democrat candidate; in another, they may adjust the vote among Hispanic individuals toward the Republican candidate. 
By replicating this process many times for a wide range of characteristics, FiveThirtyEight produces outcomes that tend to be correlated in states with similar demographics} \cite{538methodPres}.

In the academic literature, many statistical models (e.g., see \cite{klarner_2008, HUMMEL2014123,Lauderdale}) combine a variety of parameters --- including state-level economic indicators, approval ratings, and incumbency --- to forecast elections. {See \cite{Lewis-Beck} for a review}. Although some of these methods \cite{klarner_2008,Campbell} blend polls and fundamental data, Abramowitz's Time for Change model \cite{abramowitz_2016} and the work of Hummel and Rothschild \cite{HUMMEL2014123} rely on fundamental data without using any polls. {Models that are based on fundamental data alone can provide early forecasts, as they do not need to wait for polling data to become available. However, these forecasts are not dynamic and do not measure current opinion. To provide both election-day forecasts and estimates of current opinion, Linzer \cite{Linzer} augmented fundamental data with recent polls using a Bayesian approach. Although the media often stresses daily variance in polls as election campaigns unfold, the political-science community has cautioned that such fluctuations are typically insignificant and may represent differences in technique between polling sources, rather than true shifts in opinion \cite{Gelman, Wlezien,Jackman}. Therefore, to account for nonrepresentative poll samples or so-called  {``house effects'' (i.e., bias that is introduced by the specific methods that each polling organization, which is often called a ``house", uses to collect and weight raw poll responses} \cite{Jackman}), some statistical models \cite{Xbox,Jackman} adjust and weight polling data in different ways (in a similar vein to FiveThirtyEight \cite{538methodPres}). One can also consider simpler alternative approaches, such as poll aggregation, for reducing error and improving the accuracy of forecasts \cite{Wang}.

Although there is extensive work on mathematical modeling of political behavior (e.g., see \cite{PhysRevLett.112.158701,BottcherPLOS,GALAM200456,Braha}) and opinion models more generally \cite{castellano09,porter2016}, most such studies {focus} on phenomena such as opinion dynamics or on questions that are related only tangentially to elections, rather than on engaging with data-driven forecasting. For example, Braha and de Aguiar \cite{Braha} and Fern\'{a}ndez-Gracia \emph{et al}. \cite{PhysRevLett.112.158701} combined generalized voter models with data on election results to comment on vote-share distributions and correlations across U.S. counties. In a series of papers (e.g., \cite{GalamTrump,GalamUnavow}), Galam used a ``sociophysics'' approach (without reliance on polls or fundamental data) to suggest race outcomes and shed light on the dynamics that may underlie various election results. {Very recently, Top\^{i}rceanu} \cite{Topirceanu} {developed a temporal attenuation model for U.S. elections with a basis in national polling data. This work} \cite{Topirceanu} {includes measurements of each candidate's momentum in time and is able to produce forecasts at the national level.}

Accounting for interactions between states is crucial for producing reliable {forecasts, and FiveThirtyEight's Nate Silver }\cite{538methodPres} has stressed the importance of correlating polling errors by state demographics. Such correlations play an important role in the forecasts of FiveThirtyEight. Their approach \cite{538methodPres}, which one can view as indirectly incorporating relationships between states through noise, relies on {geographic closeness or demographic similarity between states}; these are inherently undirected quantities. For example, if Ohio and Pennsylvania are viewed as similar by FiveThirtyEight, so are Pennsylvania and Ohio. However, it is possible that states influence each other in directional ways. For example, voters in Ohio may more strongly influence the population in Pennsylvania than vice versa. {The strength with which states influence each other can depend on where candidates are campaigning, the people with whom voters in different states interact, which distant states are featured prominently in the news, which states most resonate with local voters, and other factors.} {Linzer \cite{Linzer} estimated national-level influences on state voters on a daily basis using a statistical modeling approach, but we are not aware of prior work that has estimated directed, asymmetric state--state relationships. {We are also not aware of} poll-based forecasting approaches that take a mathematical modeling perspective.}

To make election forecasting more transparent, {broaden the community that engages with polling data,} and raise research questions {from a dynamical-systems perspective}, we propose a data-driven mathematical model of the temporal evolution of political opinions during U.S. elections. We use a poll-based, poll-aggregating approach to specify model parameters, allowing us to provide quantitative forecasts of the vote margin by state. {We consider simplicity in the election-specific components of our model} a strength and thus do not weight or adjust the polling data in any way. Following Wang \cite{Wang}, we strive to be fully transparent; we provide all of our code, data, and detailed reproducibility instructions in our GitLab repository \cite{Gitlab_elections}. We have a special interest in exploring how states influence each other, and (because it provides a well-established way to frame such asymmetric relationships) we borrow techniques from the field of disease modeling. Using a compartmental model of disease dynamics, we treat Democrat and Republican voting intensions as contagions that spread between states. Our model performs well at forecasting the 2012 and 2016 races; and we use it to forecast the 6~November 2018 U.S. gubernatorial and senatorial elections. {We posted our forecasts \cite{preprint} on arXiv on 5~November 2018.} {For the 2018 senatorial races, we also explore how early we can make accurate forecasts, and we find that our model is able to produce stable forecasts from early August onward.} 

{Admittedly, our forecasting method involves many simplifications; our goal is to apply a data-driven, dynamical-systems approach to elections that we hope leads to more such studies in the future.} Most importantly, our model demonstrates how one can employ mathematical tools (e.g., dynamical systems, uncertainty quantification, and network analysis) to help demystify forecasting, explore how subjectivity and uncertainty impact forecasting, and suggest future research directions in the study of political elections. See \cite{2020forecasts} for our model's forecasts for the 2020 U.S. elections, which are forthcoming at the time of this writing.

%%%%%

 \section{{Background:} Compartmental modeling of infections}

   \begin{figure}[t!]
 \centering
 \includegraphics[width=0.8\linewidth]{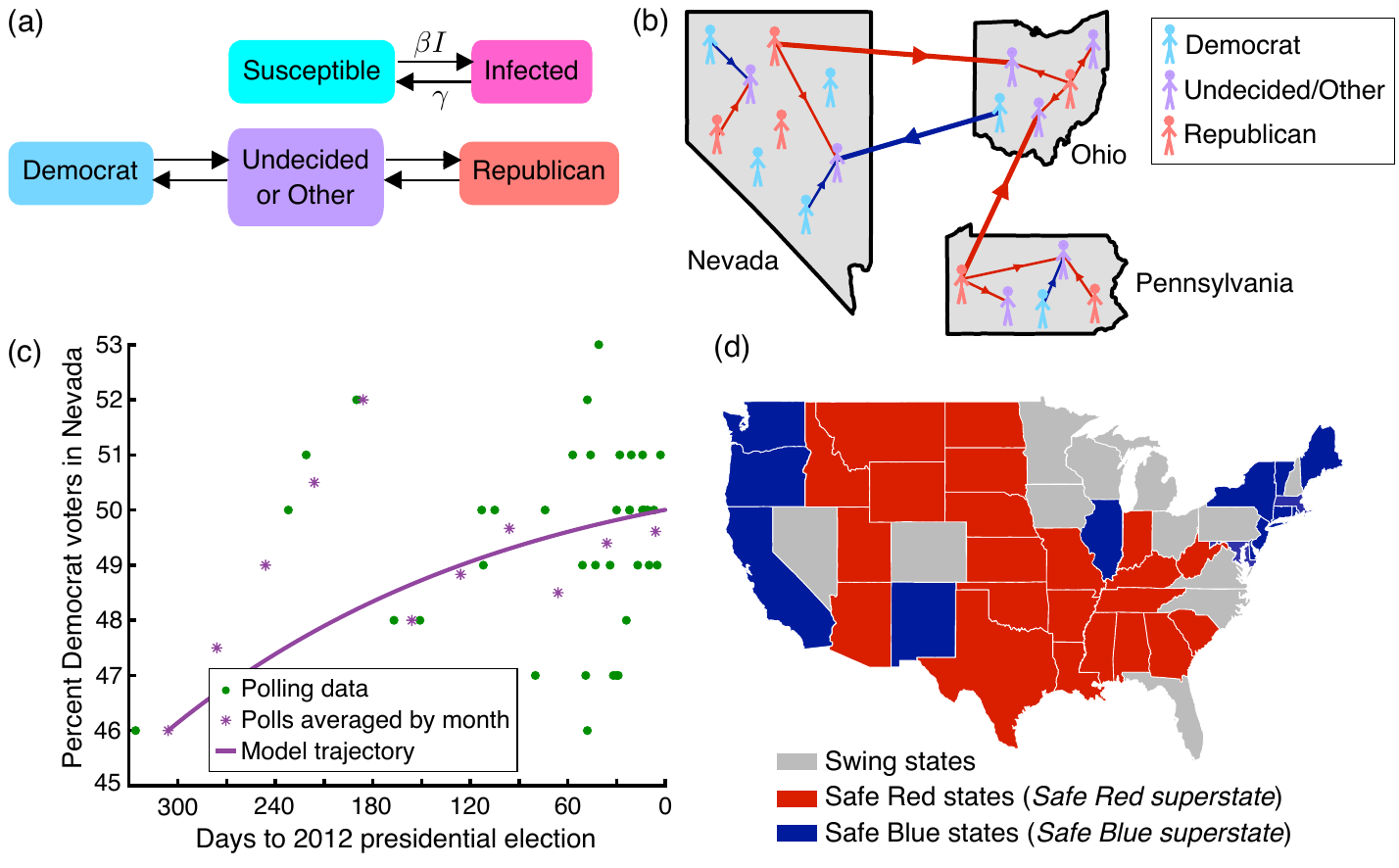}
 \caption{Overview of our modeling approach. (a) A  susceptible--infected--susceptible (SIS) compartmental model tracks the {fraction} of susceptible ($S(t)$) and infected individuals ($I(t)$) in a community in time; these quantities evolve according to infection and recovery. We repurpose this approach by including two types of infections (Democrat and Republican voting inclinations), interpreting infection as opinion adoption, and replacing recovery with turnover of committed voters to undecided ones. (b) We assume that individuals interact within and between states. {Red (respectively, blue) lines indicate interactions between undecided voters and Republicans (respectively, Democrats), and thick and thin lines correspond to interactions between voters in different states and in the same state, respectively.} In this cartoon example, Pennsylvania influences Ohio, but voters in Nevada and Pennsylvania do not interact. {We show individual voters and their interactions in this figure, but our model works at a population level and tracks voter percentages. Additionally, this image is a schematic and is not to scale.} (c) Example model dynamics: Democrat voting inclination in Ohio leading up to the 2012 presidential election. {We take the mean of the polls by month to obtain the data points that we show as purple asterisks. We specify parameters by minimizing the difference between our model \eqref{eq:elec1}--\eqref{eq:elec3} and these monthly data points.} 
We simulate the temporal evolution of opinions in the year leading up to each election, but we focus on the result at time $t=0$ days to the election. (d) For elections with many state races, we combine all reliable Republican and Democrat regions into two ``superstates'' (in Red and Blue, as we use traditional party colors). We show the superstates for presidential elections; see Table \ref{table:s1} for the superstates that we use in other elections. 
 }
 \label{fig:overview}
 \end{figure}

 {Because we repurpose our forecasting approach from {compartmental} modeling {of disease dynamics} \cite{HethcoteReview,Diekmann,ccc}, we begin with an introduction to these techniques. Compartmental models \cite{Kermack700,Kermack55,Kermack94,HethcoteReview,Diekmann,ccc} are a standard mathematical approach for studying biological contagions, {including the current COVID-19 pandemic }\cite{covid}. Developed initially for analyzing the spread of diseases such as influenza \cite{Coburn2009}, compartmental models (which are often combined with network structure to incorporate social connectivity \cite{kiss2017,pastor-Satorras2015}) have also been used to study phenomena such as social contagions \cite{FrenchRiots, IdeaSpread}. Compartmental models are built on the idea that one can categorize individuals into a few distinct types (i.e., ``compartments''). One then describes contagion dynamics using flux terms between the various compartments. For example, a susceptible--infected--susceptible (SIS) model divides a population into two classes. At a given time, an individual can be either susceptible (S) or infected (I). As we illustrate in Figure~\ref{fig:overview}a, the fraction of susceptible and infected individuals in a community depends on two factors:
\begin{itemize}
\item[$\bullet$] \emph{transmission}: when an infected individual interacts with a susceptible individual, the susceptible person has some chance of becoming infected; and
\item[$\bullet$] \emph{recovery}: an infected person has some chance of recovering and becoming susceptible again.
\end{itemize}

Suppose that $S(t)$ and $I(t)$, respectively, are the fraction of susceptible and infected individuals in a well-mixed population at time $t$. One then describes infection spread using the following set of ordinary differential equations (ODEs):
\begin{align} 
	\frac{dS}{dt} &= \underbrace{\gamma I}_{\text{recovery with rate $\gamma$}} - \underbrace{\beta SI}_{\text{infection with rate $\beta$}}\,, \label{eq:SIS1} \\
	\frac{dI}{dt} &= -\gamma I + \beta SI\,, \label{eq:SIS2}
\end{align}
where $S(t) + I(t) = 1$ and $\beta$ and $\gamma$ correspond to the rates of disease transmission and recovery, respectively. {We provide additional background on compartmental models in} Section \ref{compartmental}, {but it may be helpful to think of the equations $\frac{dS}{dt} = f(S,I)$ and $\frac{dI}{dt} = g(S,I)$, where $f$ and $g$ are unknown functions. If we Taylor expand the functions $f$ and $g$, then Eqns.}~\eqref{eq:SIS1}--\eqref{eq:SIS2} {are the lowest-order terms in this expansion that make sense for our application. In particular, the term for transmission must depend on both $S$ and $I$, because both susceptible and infected individuals need to be present for a new infection to occur.} SIS models have been extended to account for more realistic details, such as multiple contagions, communities, and contact structure between individuals or subpopulations \cite{kiss2017,pastor-Satorras2015,Allen2diseases,McCormackMultiPatch}.

%%%%%% 
\section{Model and methods}
\label{sec:methods}
We now construct a model of party choices in elections in the form of a compartmental model, {describe how we specify its parameters, and overview how we incorporate uncertainty into our forecasts. See} Section \ref{sec:summary} {and Figure}~\ref{fig:algorithm} {for a summary of the steps that we follow to generate a forecast; we also summarize some of the main simplifications that our method involves in} Section \ref{sec:summary}.} Our code and the data sets that we use in our model are available on GitLab \cite{Gitlab_elections}.

%%%%%%
\subsection{Our election model}
Our model for election dynamics is a two-pronged SIS compartmental model (see Figure~\ref{fig:overview}a). First, we reinterpret ``susceptible'' individuals as undecided (or independent or minor-party) voters. Because most U.S. elections are dominated by two parties, we consider two contagions: Democrat and Republican voting intentions. We track these quantities within each state and make the assumption that populations are well-mixed within each state. Let
\begin{align*} 
	S^i &= \text{fraction of undecided voters in state $i$}\,, \\
	I^i_\text{D} &= \text{fraction of Democrat voters in state $i$}\,, \\
	I^i_\text{R} &= \text{fraction of Republican voters in state $i$}\,,
\end{align*}
where $I^i_\text{D} + I^i_\text{R} + S^i = 1$. We account for four behaviors:
\begin{itemize}
\item[$\bullet$] \emph{Democrat transmission}: An undecided voter can decide to vote for a Democrat due to interactions with Democrats. As we discuss below, we interpret ``interactions'' and ``transmission" broadly.
\item[$\bullet$] \emph{Republican transmission}: An undecided voter can decide to vote for a Republican due to interactions with Republicans. 
\item[$\bullet$] \emph{Democrat turnover}: An infected person has some chance of changing their mind to undecided (this amounts to ``recovering'').
\item[$\bullet$] \emph{Republican turnover}: An infected person has some chance of becoming undecided (i.e., recovering).
\end{itemize}
{We assume that committed voters can ``transmit" their opinions to undecided voters but that undecided voters do not sway Republicans or Democrats. Marvel \emph{et al.} used a similar assumption in a model of more general ideological conflict} \cite{Marvel2012}. Additionally, although the language of contagions does not necessarily apply to social dynamics \cite{LehmannSune2018}, we find it useful for our work. These terms highlight that our model is not a specialized election model, as part of our goal is to show how a general framework can give meaningful forecasts in high-dimensional systems. We thus expect similar ideas to provide insight into forecasting in many complex systems.

By extending the traditional SIS model \eqref{eq:SIS1}--\eqref{eq:SIS2} to account for two contagions and $M$ states or ``superstates'' (see Figure~\ref{fig:overview}d), we obtain the following ODEs:
\begin{align} 
	\frac{dI_\text{D}^i}{dt}(t) &= \underbrace{-\gamma^i_\text{D} I^i_\text{D}}_\text{Dem.\ loss} + \underbrace{\sum_{j=1}^{M} \beta^{ij}_\text{D} \frac{N^j}{N} S^iI^j_\text{D}}_\text{Dem.\ infection}\,, \label{eq:elec1} \\
 	\frac{dI_\text{R}^i}{dt}(t) &= \underbrace{-\gamma^i_\text{R} I^i_\text{R}}_\text{Rep.\ loss} + \underbrace{\sum_{j=1}^{M} \beta^{ij}_\text{R} \frac{N^j}{N} S^iI^j_\text{R}}_\text{Rep.\ infection}\,, \label{eq:elec2} \\
	  \frac{dS^i}{dt}(t) &= \gamma^i_\text{D} I^i_\text{D} + \gamma^i_\text{R} I^i_\text{R}  -\sum_{j=1}^{M} \beta^{ij}_\text{D} \frac{N^j}{N} S^iI^j_\text{D}- \sum_{j=1}^{M} \beta^{ij}_\text{R} \frac{N^j}{N} S^iI^j_\text{R} \,, \label{eq:elec3}
\end{align}
where $N$ is the total number of voting-age individuals in the U.S.; $N^j$ is the number of voting-age individuals in state $j$ \cite{FedRegister2016,FedRegister2017,FedRegister2012}; and $\gamma_\text{D}^i$ and $\gamma_\text{R}^i$ describe the rates of committed Democrat and Republican voters, respectively, converting to undecided. Similarly, $\beta^{ij}_\text{D}$ and $\beta^{ij}_\text{R}$ correspond, respectively, to the transmission (i.e., influence) rates from Democrat and Republican voters in state $j$ to undecided individuals in state $i$. {In addition to the presence of state-level variables and two contagions in Eqns.} \eqref{eq:elec1}--\eqref{eq:elec3}, {the other difference between our election model and the traditional SIS system in Eqns.} \eqref{eq:SIS1}--\eqref{eq:SIS2} {is the new terms that include the factor} $N^j/N$. {These terms appear in Eqns.} \eqref{eq:elec1}--\eqref{eq:elec3} {because of how we choose to approximate the mean number of interactions between undecided voters in state $i$ and committed voters in state $j$. See} Section \ref{compartmental} {for details.}

{The $\beta$ parameters, which pertain to transmission dynamics, allow us to model directed relationships as a form of network structure between states. We take a broad interpretation of ``transmission". Although opinion persuasion (i.e., transmission) can occur through communication between undecided and committed voters \cite{Pons}, we expect that it can also occur through campaigning, news coverage, and televised debates. We hypothesize that these venues are an indirect means for voters in one state to influence voters in another state. For example, if news coverage of Republican campaigning in Pennsylvania resonates with undecided voters in Ohio, there may be an associated indirect route of opinion transmission from Pennsylvania to Ohio. Therefore, we consider large $\beta^{ij}_\text{R}$ to signify that Republicans in state $j$ strongly influence undecided voters in state $i$, and such ``strong influence'' may be due either to conversations (or other direct interactions) between voters or due to indirect effects like state--state affinities that are influenced or activated by media.

\subsection{Parameter fitting} 
\label{fit}
Broadly, we obtain our parameters in Eqns. \eqref{eq:elec1}--\eqref{eq:elec3} by fitting to about a year (or less, in the case of our early forecasts) of state polls} \cite{HuffPostPollster,RealClearPoliticsData}. We gather these polls from HuffPost Pollster \cite{HuffPostPollster} for our 2012 and 2016 forecasts and from RealClearPolitics \cite{RealClearPoliticsData} for our 2018 forecasts. Our parameters are different for each election and year, as we use the data that are specific to each race for fitting. To fit model parameters for a given election and year (e.g., the 2012 senatorial races), we begin by formatting its associated polling data. First, we assign each poll a time point by taking the mean of its start and end dates. Because some states are polled more frequently than others, {we then adjust our data so that each state or superstate has an equal number of data points that we can use to fit our model.} We do this by binning the polls for each state in $30$-day increments that extend backward from the appropriate election day (6~November 2012, 8~November 2016, or 6~November 2018). Our earliest bin includes polls from between $330$ and $300$ days before an election\footnote{We made one adjustment to this rule for the 2012 presidential race. For this race, our earliest bin includes polls from between $400$ and $300$ days until the election.}, so the maximum number of bins that we consider is $11$. {Most of our forecasts are in November; for these, we use all $T=11$ bins.} For our earlier forecasts in Figure~\ref{fig:gov1}, we use a smaller subset of the polling data. As an example, to forecast the 2018 senatorial elections on $7$ August, we bin the polling data from between $330$ and $90$ days of the election in $30$-day increments to obtain $T=8$ bins.

{Within each $30$-day bin, we take the mean of the polling data to help remove small-scale fluctuations in the polls.} If a given state has no polls within a bin, we approximate the associated data point using linear interpolation. In many cases, there are no polls for a state early in the year, so we set all missing early data points for that state to its earliest data point. To arrive at $T$ data points each for the Safe Red and Safe Blue superstates, for each of these $T$ points, we take a weighted average of the individual data points of each of the states within these conglomerates. To determine the weightings, we use the states' voting-age population sizes. {The result of this process is $T$ data points per state or superstate that we forecast; see Figure}~\ref{fig:overview}c {for an example.}

{Let $\{R^j(t_i), D^j(t_i), U^j(t_i)\}_{i=1,\ldots, T}$ denote the $T$ data points for state (or superstate) $j$ that we obtain through the above process. The variables $R^j(t_i)$, $D^j(t_i)$, and $U^j(t_i)$ are the fractions of Republicans, Democrats, and others in state $j$ at time point $t_i$. To describe our fitting procedure, we define a ``concentration'' vector 
\begin{align*}
	\textbf{C}(t_i) &= [R^1(t_i),\ldots,R^M(t_i),D^1(t_i),\ldots,D^M(t_i),U^1(t_i),\ldots,U^M(i)]\,,
\end{align*} 
where $M$ is the number of states and superstates that we forecast for a given election (see Table~\ref{table:s1}). For a candidate parameter set, $({\beta},{\gamma}) = \{{\beta}^{jk}_\text{R},{\beta}^{jk}_\text{D},{\gamma}^j_\text{R},{\gamma}^j_\text{D}\}_{j,k=1,\ldots,M}$, we define $\textbf{c}^{\beta,\gamma}$ to be the solution of Eqns.~\eqref{eq:elec1}--\eqref{eq:elec3} using these parameters:
\begin{align*}
	\textbf{c}^{{\beta},{\gamma}}(t) &= [I^1_\text{R}(t),\ldots,I^M_\text{R}(t),I_\text{D}^1(t),\ldots,I_\text{D}^M(t),S^1(t).\ldots,S^M(t)]\,.
\end{align*} 
We obtain our parameters $(\hat{\beta}, \hat{\gamma})$ for a given election by minimizing the least-squares deviation between the averaged polling data and the solutions of Eqns.~\eqref{eq:elec1}--\eqref{eq:elec3} at the $T$ time points. That is,
\begin{align*} 
	(\hat{\beta},\hat{\gamma})=\underset{({\beta},{\gamma})}{\mathrm{argmin}} \sum_{i=1}^{T} \|\textbf{C}(t_i)-\textbf{c}^{{{\beta},{\gamma}}}(t_i)\|_2^2\,.
\end{align*}  
These parameter estimates are consistent and converge weakly to a Gaussian distribution if the data is from a density-dependent Markov jump process \cite{rempala2012least}.}

{
We base our parameters for a given election on $2 \times T \times M$ data points that represent the percentages of Republican and Democrat voters at $T$ time points in each of $M$ states or superstates. (We also know the percentages of other voters, but these data points are correlated with the above data, because $R^j(t_i) + D^j(t_i) + U^j(t_i)=1$.) As a comparison, there are $2\times M$ turnover parameters and $2 \times M^2$ transmission parameters in Eqns.} \eqref{eq:elec1}--\eqref{eq:elec3}. {For example, we use $308$ data points to specify $420$ parameters in our November forecasts of presidential elections, which include $M=14$ states and superstates.}

Our GitLab repository \cite{Gitlab_elections} {includes all of the model parameters that we generate using the procedure that we just described. Across all of the election years (2012, 2016, and 2018) and election types (gubernatorial, senatorial, and presidential) that we consider, the mean minimum transmission parameter is $0.000000$ and the mean maximum transmission parameter is $0.555818$. Our recovery parameters $\{\gamma_\text{R}^i, \gamma_\text{D}^i\}$ vary from a mean minimum of $0.000000$ to a mean maximum of $0.047622$. As an example, we illustrate the parameters for the 2018 senatorial races in Figures}~\ref{fig:par1}--\ref{fig:par3}.

%%%%

\subsection{Uncertainty} \label{sec:SDEmodel}
{For some of our forecasts (specifically, for our 2018 forecasts and for our case study of the 2016 presidential race in} Section \ref{sec:trump}), {we generalize our model} \eqref{eq:elec1}--\eqref{eq:elec3} to a system of stochastic differential equations (SDEs):
\begin{align} 
 {dI_\text{D}^i}(t) &= \underbrace{\left(-\gamma^i_\text{D} I^i_\text{D}+\sum_{j=1}^{M} \beta^{ij}_\text{D} \frac{N^j}{N} S^iI^j_\text{D}\right)}_\text{deterministic dynamics in Eqn.~\eqref{eq:elec1}}dt + \underbrace{\sigma dW^i_\text{D}(t)}_{\text{uncertainty}}\,, \label{eq:sde1} \\
 	{dI_\text{R}^i}(t) &= \left(-\gamma^i_\text{R} I^i_\text{R} + \sum_{j=1}^{M} \beta^{ij}_\text{R} \frac{N^j}{N} S^iI^j_\text{R}\right)dt + \sigma dW^i_\text{R}(t)\,, \label{eq:sde2} \\
	  {dS^i}(t) &= \left(\gamma^i_\text{D} I^i_\text{D} + \gamma^i_\text{R} I^i_\text{R}  -\sum_{j=1}^{M} \beta^{ij}_\text{D} \frac{N^j}{N} S^iI^j_\text{D} - \sum_{j=1}^{M} \beta^{ij}_\text{R} \frac{N^j}{N} S^iI^j_\text{R}\right)dt + \sigma dW^i_\text{S}(t)\,,   \label{eq:sde3}
\end{align}
where we now consider $I_\text{D}^i, I_\text{R}^i,$ and $S^i$ to be stochastic processes and let $W^i_\text{D},W^i_\text{R}$, and $W^i_\text{S}$ be Wiener processes. {The parameters in this system have the same values as those that we fit for the corresponding deterministic model} \eqref{eq:elec1}--\eqref{eq:elec3}. By simulating many (e.g., we use $10,000$) elections using Eqns.~\eqref{eq:sde1}--\eqref{eq:sde3}, we obtain a distribution of possible outcomes; this allows us to quantify uncertainty in a given race. {We explore the effects of uncorrelated and correlated noise on our forecasts in} Section \ref{sec:trump}.

{In graphics that describe its 2016 presidential forecasts, FiveThirtyEight} \cite{538pres2016forecast} {included both its expected vote margin for each state and a confidence interval that indicates the range in which the middle $80$\% of its model outcomes fell for that state. To determine our noise strength $\sigma$ in Eqns.} \eqref{eq:sde1}--\eqref{eq:sde3}{, we measure the length of these confidence intervals in FiveThirtyEight's final forecasts. Based on our estimates, the intervals range in length from about $13$ to $19$ percentage points for swing states. We tested a few values of $\sigma$ and chose $\sigma = 0.0015$ to roughly match the length of our $80$\% confidence intervals}\footnote{{The middle $80$\% confidence interval is the range in which a set of outcomes lies after we have removed the bottom $10$\% and top $10$\% of outcomes from the set. We determine these cutoffs using the built-in} {\sc prctile} {function in} {\sc Matlab} (version 9.3).}
{for the vote margin to these measurements for our 2016 presidential forecasts. For example, across the $10,000$ simulations of Eqns.} \eqref{eq:sde1}--\eqref{eq:sde3} {that we show in Figure}~\ref{fig:uncertaintyTrump}b, {the mean length of our $80$\% confidence intervals for swing states is about $15$ percentage points.}

\subsection{Summary of our approach and important simplifications}
\label{sec:summary}

\begin{figure}[t!]
\centering
\includegraphics[width=0.8\textwidth]{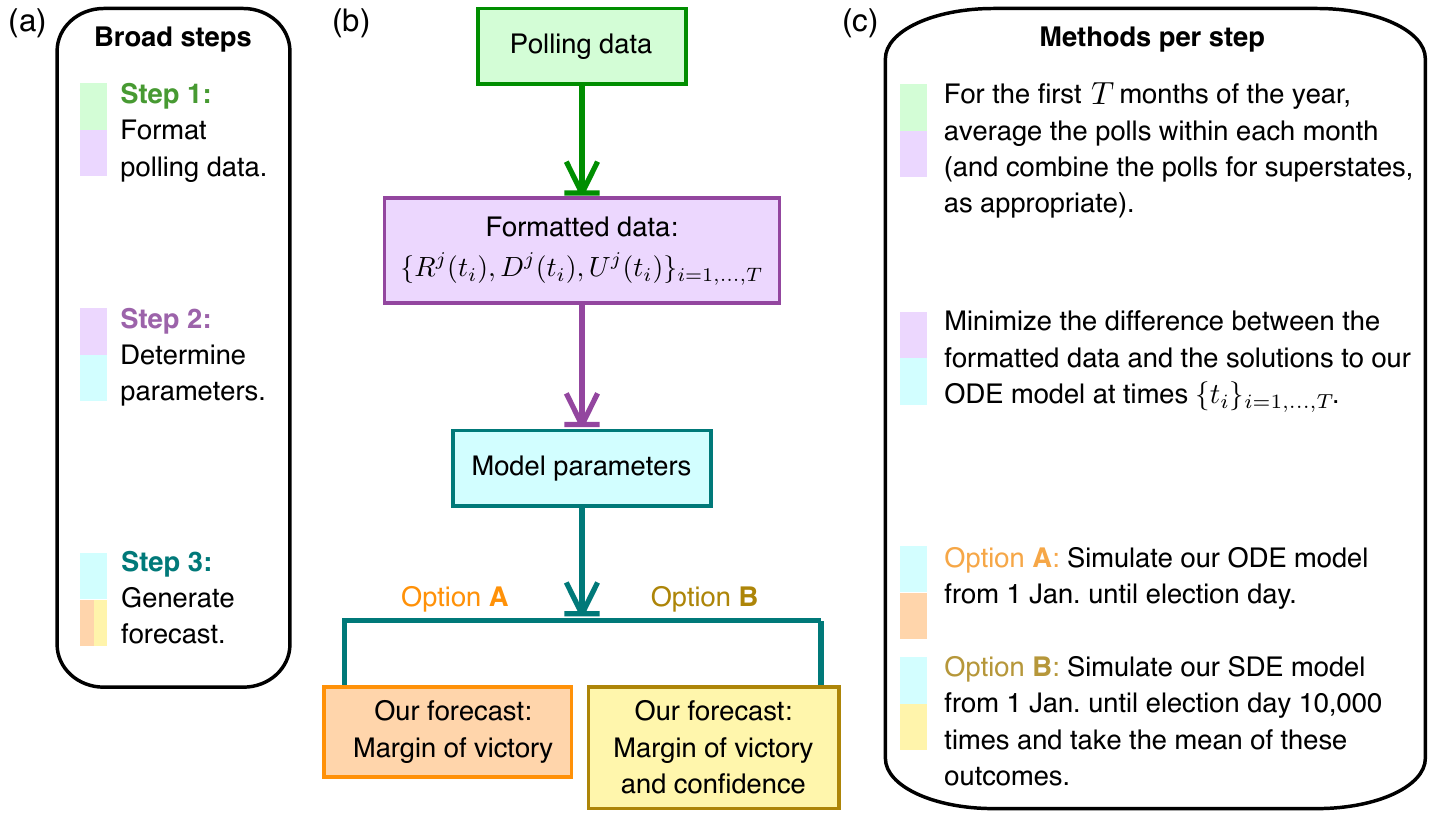}
\caption{{Summary of our approach. (a) To forecast a given election, we follow three steps. We start by formatting polling data from} \cite{HuffPostPollster,RealClearPoliticsData}. 
{(b) Once we fit our model parameters, we have two options for forecasting elections. When using Option A, we generate forecasts of the vote margin by simulating our ODE model} \eqref{eq:elec1}--\eqref{eq:elec3} {from 1 January through election day. When using Option B, we instead simulate many realizations of our SDE model} \eqref{eq:sde1}--\eqref{eq:sde3} {to produce a distribution of vote margins, and we take the mean of this distribution to be our forecast vote margin.} {(c) Our final forecasts in November have $T=11$ months, bur earlier forecasts for the 2018 races (see Figure}~\ref{fig:gov1}) {have $T <11$ months}.  
 }\label{fig:algorithm}
\end{figure}

{Our forecasting process consists of three steps, which we illustrate in Figure}~\ref{fig:algorithm}. {Our first step is to take the mean of the polling data} \cite{HuffPostPollster,RealClearPoliticsData} {within each month (specifically, within each 30-day increment extending backward from election day) to generate $T$ data points in time for each state or superstate.} {Because we focus on November forecasts, there are $T=11$ months in most of our simulations. This corresponds to one data point in each month from January to November of an election year. For our earlier forecasts for the 2018 races in Figure}~\ref{fig:gov1}, there are {$T <11$ months. As an example, our forecasts in August have $T=8$. Second, we fit our parameters to this trajectory of polling data in the first $T$ months of a year using our deterministic model} \eqref{eq:elec1}--\eqref{eq:elec3}. 
Third, after we fit our parameters to polls for a given election, we use one of our models to simulate the daily evolution of political opinions from the preceding January through election day. In such simulations, we use the polling data point at $T=1$ to specify our initial conditions (see Section \ref{numerical}). {Because we have both a deterministic model}~\eqref{eq:elec1}--\eqref{eq:elec3} {and a stochastic model}~\eqref{eq:sde1}--\eqref{eq:sde3}, {we have two options for simulating elections. For our initial study in} Section \ref{sec:initial}, {we use our ODE model}~\eqref{eq:elec1}--\eqref{eq:elec3} {to simulate elections. (We label this approach as Option A in Figure}~\ref{fig:algorithm}.) {Following this initial study, we instead generate forecasts by simulating many realizations of our SDE model}~\eqref{eq:sde1}--\eqref{eq:sde3}. {This alternative approach (which we label as Option B in Figure}~\ref{fig:algorithm}) {provides distributions of the vote margin, allowing us to forecast the margin of victory (namely, the mean of this distribution) in each state and our confidence in this margin.} Except for Figures~\ref{fig:senateSI2}b,c, we use a time step of $\Delta t = 3$ days for parameter fitting, specify a time step of $\Delta t = 0.1$ days for simulating our models once we have determined their parameters, and simulate $10,000$ realizations of Eqns.~\eqref{eq:sde1}--\eqref{eq:sde3} to generate the forecasts that are based on our SDE model. We refer to these values as our ``typical" simulation parameters\footnote{Our forecasts in Figures~\ref{fig:senateSI2}b,c are the only situations in which we do not use our typical simulation parameters. We determined the model parameters for our simulations in Figures~\ref{fig:senateSI2}b,c using a time step of $\Delta t = 15$ days, and we then produced our forecasts based on $4,000$ simulations of Eqns. \eqref{eq:sde1}--\eqref{eq:sde3} with a time step of $\Delta t = 0.1$ days.}. In each figure caption, we indicate whether we use our ODE model or our SDE model for the simulations in that figure.

We do not claim that our approach is the most accurate method of forecasting elections. Instead, we propose it as a data-driven model that admittedly involves many simplifications, some of which are instructive to mention before we discuss our simulation results. Important simplifications include the following:
\begin{itemize}
\item[$\bullet$] {Although it is generally not realistic, we assume that voters mix uniformly (e.g., everyone has the same influence on everyone else), aside from the state structure, which is analogous to {patches} in epidemiology; accounting for {additional network structure} may improve forecasts \cite{Meyers, Watts,Busenberg}.}
\item[$\bullet$] We combine all sources of opinion adoption into time-independent transmission parameters $\beta^{ij}_\text{R}$ and $\beta^{ij}_\text{D}$. {Including time-dependent transmission parameters may lead to richer model behavior, such as oscillations over time in a state's vote margin.}
\item[$\bullet$] {If undecided voters remain at the end of our simulation, we assume that they vote for minor-party or other candidates.} 
\item[$\bullet$] {We assume that all polls are equally accurate. Unlike FiveThirtyEight \cite{538methodPres}, we do not weight polls more strongly based on recency or make any distinction between partisan and non-partisan polls (or polls of likely voters, registered voters, or all adults). Notably, Wang \cite{Wang} has illustrated that, when aggregated, polling data may not need to be weighted or adjusted to account for polling source to be accurate.} 
\item[$\bullet$] {Because of our approach to fitting parameters, we use the earliest available formatted polling data to initialize our models. Using data-assimilation techniques} \cite{Stuart} {to determine initial conditions may improve the accuracy of our forecasts.}
\end{itemize}
Despite these simplifications, our forecasting method performs as well as popular analysts. We discuss these results in Section \ref{sec:main}.

%%%%%%%%
\section{Results}
\label{sec:main}
We now use our ODE model \eqref{eq:elec1}--\eqref{eq:elec3} to simulate past races for governor, senator, and president in 2012 and 2016. Because realistic forecasts should incorporate uncertainty, we follow this exploration of past races with a short study of the impact of noise on our 2016 presidential forecast. To do this, we compare simulations of our SDE model \eqref{eq:sde1}--\eqref{eq:sde3} with uncorrelated and correlated noise. We conclude by using our SDE model to forecast the gubernatorial and senatorial midterms on 6~November 2018.

%%%%%%
\subsection{2012 and 2016 election forecasts}\label{sec:initial}

 \begin{figure}[t!]
 \centering
 \includegraphics[width=0.9\textwidth]{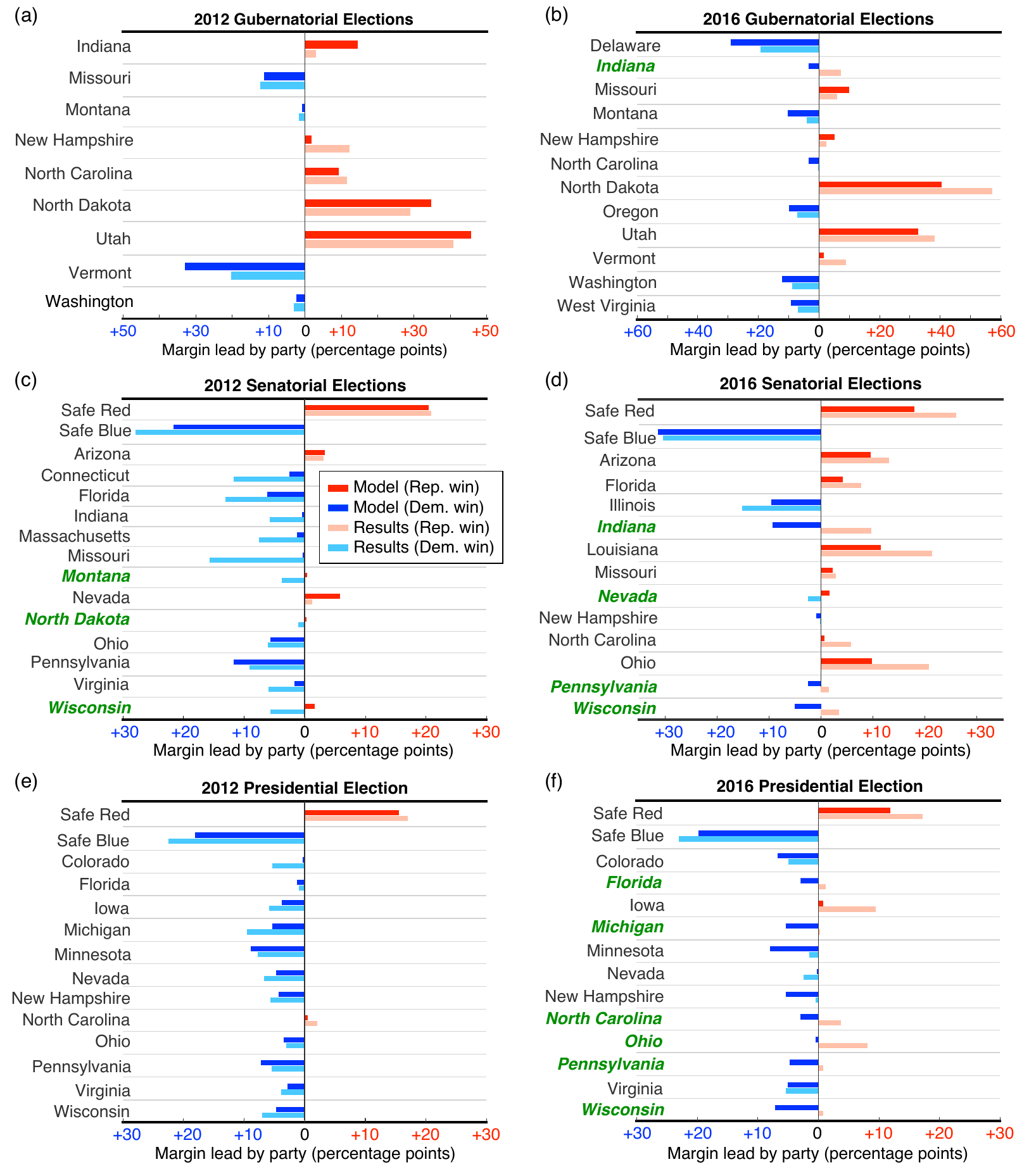}
 \caption{
{Simulations of Eqns.~\eqref{eq:elec1}--\eqref{eq:elec3}.} We calculate our 2012 and 2016 forecasts using polling data up until election day. Our forecasts do not include any election results, and they should be interpreted as forecasts from the night before an election. Comparison of our forecasts for (a, b) gubernatorial, (c, d) senatorial, and (e, f) presidential elections with results from \cite{Ballotpedia,Leip}. The horizontal axis shows the percentage-point lead by Democrats (blue) or Republicans (red). Shorter bars represent 
closer elections, and bars that extend to the right (respectively, left) correspond to Republican (respectively, Democrat) leads. The states that we forecast incorrectly are in a bold, italic green font. ``Safe Red'' and ``Safe Blue'' refer to superstates that are composed, respectively, of reliably Republican and Democrat states. We assemble the superstates based on forecaster opinions and historical data (see Section \ref{super}).\label{fig:norandom} }
 \end{figure}

By fitting our parameters to polling data for senatorial, gubernatorial, and presidential races in 2012 and 2016 without incorporating the final election results, we can simulate forecasts as if we made them on the eve of the respective election days. In Figure~\ref{fig:norandom}, we summarize our forecasts for these races. {To measure the accuracy of our forecasts, we compute a success rate at predicting (``calling'') party outcomes at the state level:}
\begin{align}
    \text{success rate} &= 100 \times \frac{\text{number of state or district races that are called correctly}}{\text{total number of state or district races that are forecast}}\,,\label{eq:success}
\end{align}
{where we consider only state and district races for which our model provides forecasts}\footnote{{As we discuss in} Section \ref{special}{, there are a few special cases in which we do not forecast a given state (see Table} \ref{table:s1}). {For example, we do not forecast single-party state races or 2012 gubernatorial races in states for which we have no polling data. See} Section \ref{alternative} {for a discussion of alternative choices that we could have made when computing accuracy in the face of these special cases.}}. As we show in Table~\ref{table:success}, our model has a similar success rate as popular forecasters 
FiveThirtyEight \cite{538pres2016forecast} and Sabato's Crystal Ball \cite{Sabato}. For example, our {success rate across all $102$ of the (state or Washington, D.C.) forecasts that we made} for presidential elections in 2012 and 2016 is $94.1\%$, whereas FiveThirtyEight and Sabato's Crystal Ball achieved success rates of $95.1\%$ and $93.1\%$, respectively.

 \begin{table}[t]
\centering
\caption{Comparison of the success rates for our model \eqref{eq:elec1}--\eqref{eq:elec3} and two popular sources. {We measure a forecaster's ``success rate" in Eqn.} \eqref{eq:success} {as the percent of (state or Washington, D.C.) races that they correctly forecast, in the sense that they identified the true winner by party, among the races that we forecast using our model (see Table} \ref{table:s1}). {Importantly, we leave races that we do not forecast (e.g., single-party races) out of these computations. See} Section \ref{special} and Section \ref{alternative} {for more details.} For FiveThirtyEight \cite{538pres2016forecast}, we use the 2016 polls-only forecast.} \label{table:success}
\begin{tabular}{lrrr}
Election & FiveThirtyEight \cite{538pres2016forecast} &Our model & Sabato \cite{Sabato} \\
\midrule
2016 presidential & 90.2\% &	88.2\% & 90.2\% \\
2016 senatorial & 90.9\% &	87.9\%  &	93.9\% \\ 
2016 gubernatorial & NA & 91.7\%	& 83.3\% \\
2012 presidential & 100\% & 100\% & 96.1\% \\
2012 senatorial & NA & 90.3\% &	93.5\%  \\
2012 gubernatorial & NA & 100\% & 77.8\% \\
\bottomrule
\end{tabular}
\end{table}

Figure \ref{fig:norandom} and Table \ref{table:success} highlight two forecasting goals: (1) estimating the vote share by state (e.g., the percents of the state vote that are received by Democrat and Republican candidates) and (2) calling the winning party by state (i.e., which party's candidate wins the election in a given state). Many qualitative forecasters (e.g., \cite{cookGov,Sabato}) focus on the second goal, whereas our model and FiveThirtyEight \cite{538govSen} pursue both goals.

%%%%
\subsection{Accounting for and interpreting uncertainty}\label{sec:trump}
{Sources of uncertainty and error in election forecasting include sampling error and systematic bias from the specific methods of different polling sources \cite{Linzer,Jackman,538methodPres,Wang}.} 
Consequently, election forecasting involves not only calling a race for a specific party and estimating vote shares, but also specifying the likelihood of different outcomes. This raises a third goal of forecasters: (3) quantifying uncertainty, such as by estimating a given candidate's chance of winning an election or by providing a confidence interval for their vote margin. We suggest that this is one of the key places where mathematical techniques can contribute to election forecasting. As a case study, we investigate two methods for accounting for uncertainty. Specifically, we compare the forecasts that result from simulating $10,000$ realizations of our SDE model~\eqref{eq:sde1}--\eqref{eq:sde3} (see Section \ref{sec:SDEmodel}) for the 2016 presidential race with uncorrelated noise with those that arise from following the same process with correlated noise.

Although U.S. elections are decided at the level of states, polling errors are correlated in regions with similar populations \cite{538methodPres}. Therefore, if a pollster is wrong in Minnesota, they may also be off in states (such as Wisconsin) with shared features \cite{538methodPres}. This type of error makes it possible for polls of a bloc of states to all be wrong together, leading to an unforeseen upset. To explore these dynamics, we compare the impact of uncorrelated noise with the effect of additive noise that is correlated on a few sample demographics. Specifically, we consider the fractions of Black, Hispanic, and adult college-educated individuals in a population. We correlate on these demographics because these data are readily available; future work should incorporate additional data.

To correlate noise in Eqns.~\eqref{eq:sde1}--\eqref{eq:sde3}, we first quantify the similarity of two states, $i$ and $j$, using the Jaccard index $J^{i,j} = \min\{D^i,D^j\}/\max\{D^i,D^j\}$, where $D^i$ is the fraction of a given demographic in state $i$ and $D^j$ is the fraction of that demographic in state $j$. The Jaccard index indicates the covariance for our increments of $\textbf{W}_\text{R}$ and $\textbf{W}_\text{S}$ in Eqns.~\eqref{eq:sde1}--\eqref{eq:sde3}. We define $\textbf{J}_\text{B}, \textbf{J}_\text{E},$ and $\textbf{J}_\text{H}$ to be the Jaccard indices that we find using the fractions of non-Hispanic Black individuals, adults without a college education, and Hispanic individuals, respectively\footnote{We compute the fraction of a given demographic in the Safe Red (respectively, Safe Blue) superstate by calculating the mean of the fractions in all of the states inside the Safe Red (respectively, Safe Blue) superstate. We do not weight these averages by state size. {Additionally, for presidential elections, we do not include the District of Columbia in our estimate of education level in the Safe Blue superstate, because data on education levels in Washington, D.C. are not listed on {\tt 247WallSt.com}} \cite{education}.}. We calculate $\textbf{J}_\text{B}$ and $\textbf{J}_\text{E}$ using 2016 U.S. Census Bureau \cite{census} data, and we base $\textbf{J}_\text{H}$ on data from \url{247WallSt.com} \cite{education}.
For our forecasts with correlated noise, each time that we simulate an election, we select one Jaccard index uniformly at random among $\textbf{J}_\text{B}$, $\textbf{J}_\text{E}$, and $\textbf{J}_\text{H}$ to use as our covariance. {For example, if we select $\textbf{J}_\text{B}$, then the increment $d\textbf{W}_\text{R}(t_n)$ has a multivariate normal distribution with mean $0$ and covariance $\textbf{J}_\text{B}$ at each time step in that simulation.} Consequently, for each such simulated election, at any given time, we are more likely to adjust the vote in a set of states with a similar feature (e.g., a high Hispanic population) in the same direction (e.g., in favor of the Democrat candidate) than we are to adjust the vote in these states in opposite directions.

  \begin{figure}[t!] 
 \centering
 \includegraphics[width=\textwidth]{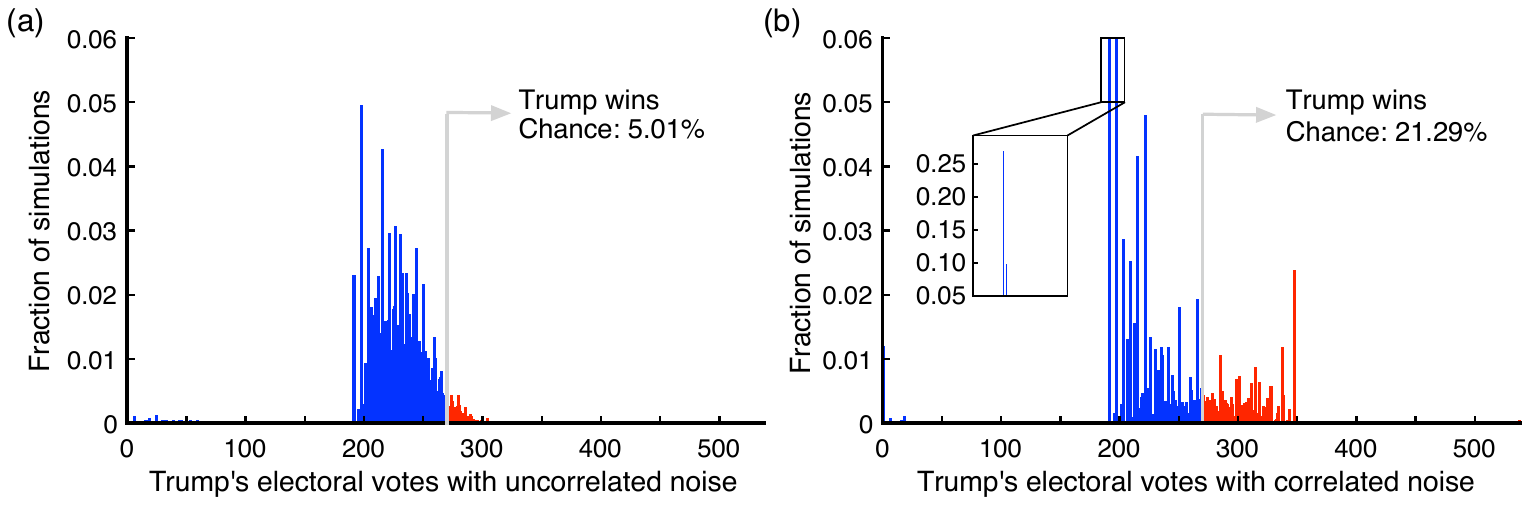}
 \caption{Impact of incorporating uncertainty in different ways, as demonstrated by our simulations of our SDE model~\eqref{eq:sde1}--\eqref{eq:sde3}, for the 2016 U.S. presidential election. (a) Uncorrelated additive white noise gives Donald Trump an approximately $5$\% chance of winning the electoral college, whereas (b) correlating noise by state demographics increases his chance of winning to about 21\%. {The tall bar in the magnified image in panel (b) illustrates that many of our simulations forecast that only the Red superstate votes Republican, leading to $191$ electoral votes for Trump. The smaller bar in the magnified image corresponds to model outcomes in which only the Red superstate and either Iowa or Nevada vote Republican.} We generate distributions by simulating $10,000$ elections using Eqns.~\eqref{eq:sde1}--\eqref{eq:sde3} with $\sigma=0.0015$.}
 \label{fig:uncertaintyTrump}
 \end{figure}

Our case study of the 2016 presidential race illustrates how accounting for uncertainty in different ways influences forecasts, echoing points that have been raised by Nate Silver and his collaborators \cite{538methodPres}. In Figure~\ref{fig:uncertaintyTrump}, we demonstrate that uncorrelated noise, which can model uncertainty in a single state or a single poll without assuming a larger systematic (e.g., country-wide) polling error, results in a low likelihood of a win by Donald Trump (the Republican candidate) in the 2016 presidential election. By contrast, correlating outcomes by demographics, which can model systematic polling errors (e.g., due to misidentifying likely voters) in similar states, increases Donald Trump's chances of winning the election by a factor of about four. This agrees with Nate Silver's comment \cite{538methodPres} that failing to account for correlated errors tends to result in underestimations of a trailing candidate's chances. {As we discuss in Section \ref{original}, because of an indexing error in one of our files, in an earlier version of our model, we correlated state outcomes on the demographics of the wrong states. After correcting this error, we obtained similar results, suggesting that it is the mere presence of correlated noise that improves Donald Trump's chances and that the noise does not need to be correlated by the specific state demographics that we used. FiveThirtyEight \cite{538methodPres}, for example, correlates state outcomes on party, region, religion, race, ethnicity, and education.}

In our computations, we do not attempt to account directly for errors in polls. Instead, we take the simple approach of assuming that we can incorporate all sources of uncertainty as an additive noise term in our SDE model \eqref{eq:sde1}--\eqref{eq:sde3}. There has been extensive work on quantifying uncertainty (see \cite{Stuart}); exploring alternative ways of accounting for uncertainty is an important future direction for research on forecasting complex systems.

%%%%%%
\subsection{2018 senatorial and gubernatorial forecasts}

 \begin{figure}[t!]
 \centering
\includegraphics[width=\textwidth]{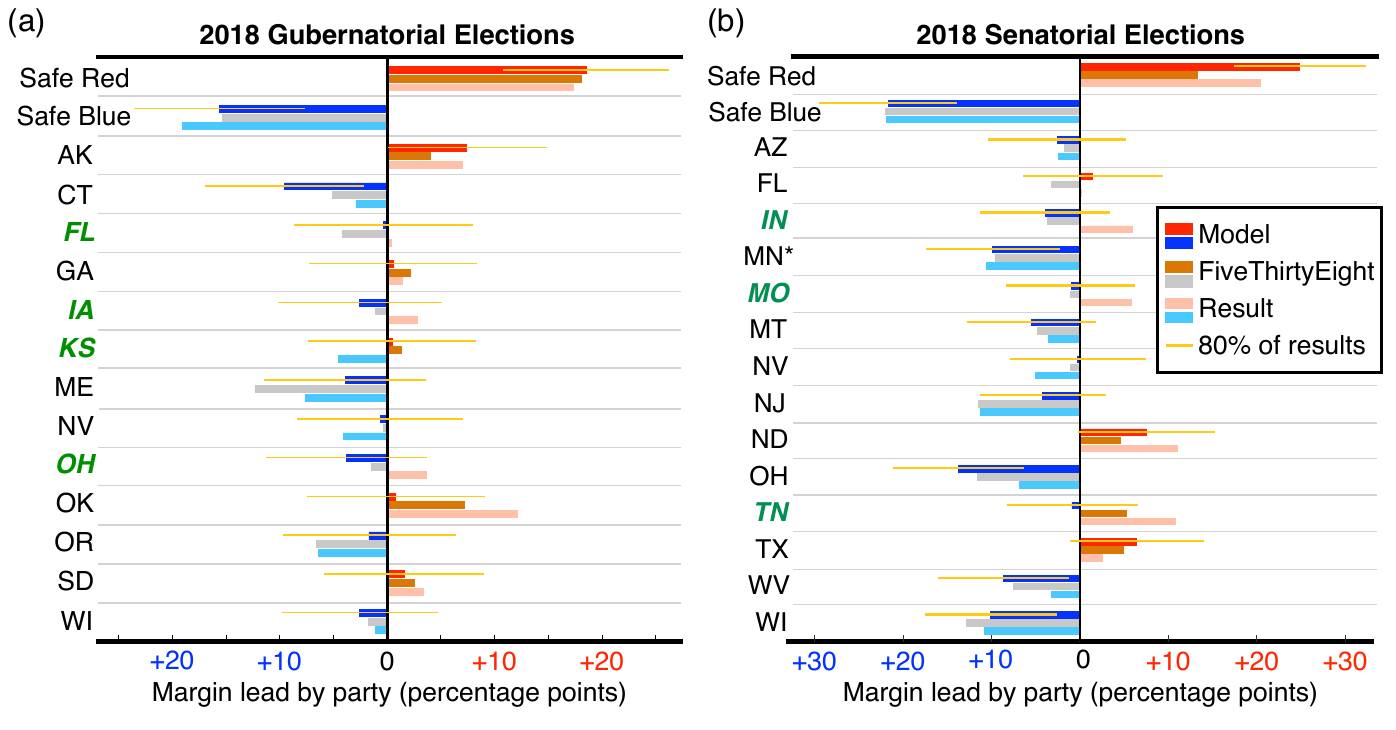}
\caption{{Forecasts of vote margins for the 2018 (a) gubernatorial and (b) senatorial races.}{ We base our forecasts on $10,000$ elections that we simulate using Eqns.~\eqref{eq:sde1}--\eqref{eq:sde3} with noise that we correlate on state demographics.} We base on forecasts on poll data that we collected from RealClearPolitics \cite{RealClearPoliticsData} through 3~November 2018. (The election took place on 6 November.) We compare them to FiveThirtyEight's 6~November forecasts (according to the ``classic'' version of the FiveThirtyEight algorithm) \cite{538govSen} and the election results \cite{NYTresults}. (We use an asterisk to mark the MN special election.) The bold, italic green font indicates state races that we called incorrectly. The length of the bars that extend to the right (in red) and left (in blue), respectively, indicate the mean percent lead by the Democrat and Republican candidate. (A value of $0$ represents a tie.) The narrow orange bars indicate the regions that encompass the middle $80$\% of our simulated election results.
 \label{fig:senateFinal}}
 \end{figure}

The 2018 midterm elections provided a fantastic opportunity for us to test our model. Our final forecasts, which we posted on 5~November (the night before the 2018 elections) \cite{preprint}, rely on polls that we gathered from RealClearPolitics \cite{RealClearPoliticsData} through 3 November\footnote{There is a time delay between when polls are completed and when the data become available from RealClearPolitics \cite{RealClearPoliticsData}. The latest polls in our gubernatorial and senatorial data sets were completed on 1~November and 2~November, respectively. Polls do not always become available {in the temporal order of polling day, so this does not imply that our data include all of the polls that occurred before these dates. For example, RealClearPolitics \cite{RealClearPoliticsData} occasionally updates its website with additional early polls, despite its prior posting of more recent polls.}} and are based on our SDE model~\eqref{eq:sde1}--\eqref{eq:sde3}. We account for uncertainty by correlating noise on education, ethnicity, and race (as in Figure~\ref{fig:uncertaintyTrump}b). Because of computational time constraints, we based our original senatorial forecast from 5~November on $4,000$ simulations of Eqns.~\eqref{eq:sde1}--\eqref{eq:sde3} and used a larger time step ($\Delta t = 15$ days) than usual for parameter fitting. Additionally, after checking our polling data without the election-day rush, we found several typos that we corrected for the forecasts in the main text. See Section \ref{original} for details. We include our original forecasts \cite{preprint} from 5~November 2018 in Figures~\ref{fig:senateSI2}--\ref{fig:govSI2}. In the main text, we present the forecasts that we obtain using our typical simulation parameters\footnote{As our typical simulation parameters, we use $\Delta t = 3$ days for parameter fitting and we simulate $10,000$ realizations of Eqns.~\eqref{eq:sde1}--\eqref{eq:sde3}. See Section \ref{sec:summary} and Section \ref{numerical} for details.}. Both forecasts project the same candidate to win in each state.

In Figures~\ref{fig:senateFinal} and \ref{fig:gov1}, we compare our gubernatorial and senatorial forecasts, respectively, for swing states and superstates with those of several popular sources. For the gubernatorial races, our Safe Red superstate consists of Alabama (AL), Arizona (AZ), Arkansas (AR), Idaho (ID), Maryland (MD), Massachusetts (MA), Nebraska (NE), New Hampshire (NH), South Carolina (SC), Tennessee (TN), Texas (TX), Vermont (VT), and Wyoming (WY); and our Safe Blue superstate consists of California (CA), Colorado (CO), Hawaii (HI), Illinois (IL), Michigan (MI), Minnesota (MN), New Mexico (NM), New York (NY), Pennsylvania (PA), and Rhode Island (RI). For the senatorial races, our Safe Red superstate consists of
Mississippi (MS), Mississippi Special (MS*), NE, Utah (UT), and WY; and our Safe Blue superstate consists of 
 Connecticut (CT), Delaware (DE), HI, Maine (ME), MD, MA, MI, MN, NM, NY, PA, RI, VT, Virginia (VA), and Washington (WA). Our results in Table~\ref{table:success2018} are based only on the remaining states, which we treat individually in our model.

Given the probabilistic nature of forecasts, it is not straightforward to evaluate their accuracy \cite{prosser_mellon_2018}, and we use the 2018 races to discuss a few ways of quantifying forecast performance. For quantitative forecasters, one natural way of evaluating performance is by computing the error in their forecasted {margin of victory (MOV)} by state. {We denote the percentages of people who voted Republican and Democrat in a given state by $R^\text{result}$ and $D^\text{result}$, respectively; and we denote the corresponding forecasted percentages $R^\text{forecast}$ and $D^\text{forecast}$. The MOV error in that 
state is}
\begin{align*}
    \text{MOV error} &= \left| \left(R^\text{results} - D^\text{results}\right) - \left(R^\text{forecast} - D^\text{forecast} \right) \right|.
\end{align*}
For example, as we showed in Figure~\ref{fig:senateFinal}a, we forecast that the Democrat candidate (Gillum) would win the Florida gubernatorial race by $0.4$ percentage points over the Republican nominee (DeSantis). DeSantis edged Gillum by $0.4$ points, so our {MOV error} is $0.8$ points for this specific race. {For close races, it is worth comparing vote margins and MOV errors to the margins of error in the polling data.} The mean margins of error that were reported in the polls \cite{RealClearPoliticsData} on which we based our parameters were $4.1$ and $4.0$ for the gubernatorial and senatorial data, respectively. Critically, this reported error is sampling error only; it does not account for other sources of error, such as ones from unrepresentative polling samples, which can result in error that is correlated by demographics \cite{prosser_mellon_2018}. {Although it is straightforward to measure accuracy using MOV error, a drawback of it is that one can use this measure only for} quantitative forecasters (see Table \ref{table:success2018}).

\begin{table}
\setlength\tabcolsep{4pt} 
\caption{Final forecast performance for the 2018 gubernatorial and senatorial races. We measure performance by calculating the mean {MOV error}, the number of state races that are missed or not called, and log-loss error. We use final forecasts, and we note that lower numbers indicate better performance. {Across these diagnostics, the forecasters perform similarly, but determining who was most successful depends on what one values in a forecast. For example, the Cook Political Report} \cite{cookGov} {identified the winning gubernatorial candidate in all of the states for which they provided a forecast, but they left $12$ states as too close to call. Our SDE model} \eqref{eq:sde1}--\eqref{eq:sde3} {provides forecasts for all of these states, but it was incorrect about $4$ of them.} The measurements in the table are for the races in Figure~\ref{fig:gov1}, and they do not include the states that we combined into the Safe Red and Safe Blue superstates. {(Additionally, see Section \ref{alternative} and Tables \ref{table:STable21} and \ref{table:STable22}.)}
}
\label{table:success2018}
\begin{tabularx}{\textwidth}{p{1.67cm} p{1cm} p{1.1cm} p{1.4cm} p{1cm} p{1cm} p{1.1cm} p{1.4cm} p{1cm}*{10}{C}}
Forecaster
& \multicolumn{4}{>{\hsize=\dimexpr4\hsize+8\tabcolsep\relax}C}{2018 Gubernatorial Races} 
& \multicolumn{4}{>{\hsize=\dimexpr4\hsize+4\tabcolsep\relax}C}{2018 Senatorial Races} \\
\cmidrule(lr){2-5} \cmidrule(lr){6-9} 
& {Mean MOV} error & Num.\ races missed & Num.\ races not called  & Log-loss error & {Mean MOV} error & Num.\ races missed & Num.\ races not called & Log-loss error \\
\midrule
Our model & 4.1\%  & 4 & 0 & 0.589 &4.6\% & 3 &0 & 0.396 \\
538 \cite{538govSen}  & 3.1\%  & 4 & 0 & 0.548&3.7\% &  3 & 0 & 0.410 \\
Sabato \cite{sabatoSenate, sabatoGov}& NA & 3 & 1 &0.585  & NA & 1 & 0 & 0.379 \\
Cook \cite{cookGov} & NA & 0 & 12 & 0.670&  NA & 0 &9 &  0.553  \\
IE \cite{insideSenGov} & NA &  2 & 3& 0.619 & NA &  1 & 1 &  0.415 \\
RCP \cite{RealClearPoliticsData} & NA & 0 & 12 &0.647 & NA & 0 & 8 & 0.565 \\
\bottomrule
\end{tabularx}
\end{table}

The baseline measure of how well forecasters do at calling race outcomes --- specifically, of whether a state will elect a Republican or a Democrat candidate --- often attracts media attention. As we illustrate in Table~\ref{table:success2018}, 2018 was a good year for a couple of the well-known pundits who use qualitative approaches, and Sabato's Crystal Ball \cite{sabatoSenate} was the most successful of these at calling outcomes. The quantitative forecasts of our SDE model \eqref{eq:sde1}--\eqref{eq:sde3} and FiveThirtyEight tied for second place (along with Inside Elections \cite{insideSenGov}). Our SDE model and FiveThirtyEight forecast the incorrect winner in the same $4$ races for governor and in $2$ of the same races for senator; we also were wrong about Tennessee, and FiveThirtyEight missed Florida \cite{538govSen}. The Cook Political Report \cite{cookGov} and RealClearPolitics \cite{RealClearPoliticsData} performed worse based on this measurement, because they left many races as toss-ups.

 \begin{figure}[t!]
\centering
\includegraphics[width=0.87\textwidth]{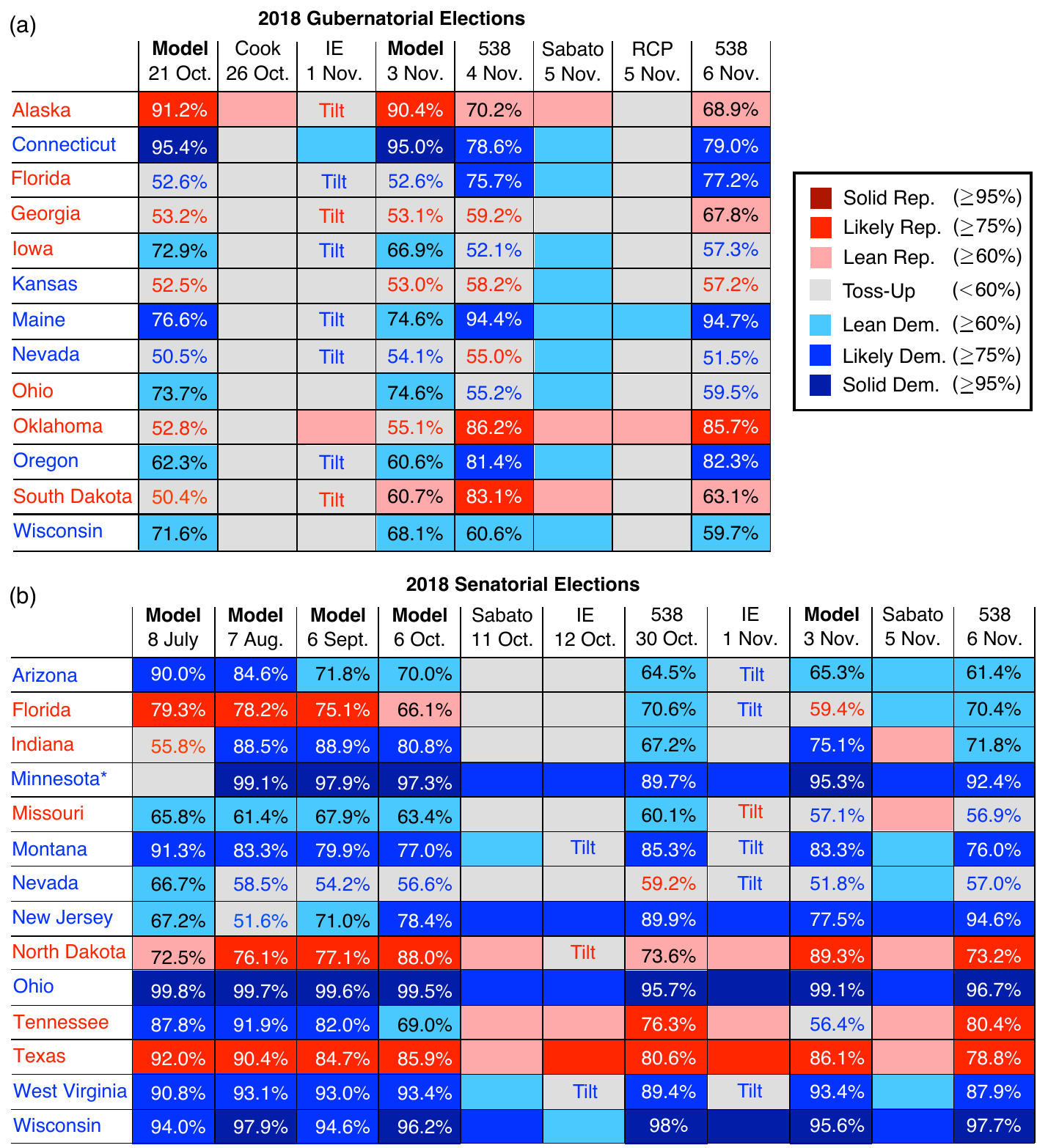}
\caption{Ratings for the 2018 (a) gubernatorial and (b) senatorial races. {There is considerable variability in these ratings across forecasters; one forecaster may identify a given race as a toss-up, and another forecaster may identify that race as solidly partisan.}
 {As we show in panel (b), our forecasts are consistent (with respect to which party we project to win a given race) after July}. {We base our forecasts on $10,000$ simulations of our SDE model~\eqref{eq:sde1}--\eqref{eq:sde3}.} We show ratings from our model, the Cook Political Report \cite{cookGov}, Inside Elections (IE) \cite{insideSenGov}, Sabato's Crystal Ball \cite{sabatoGov}, RealClearPolitics \cite{RealClearPoliticsData}, and FiveThirtyEight \cite{538govSen}. IE \cite{insideSenGov} breaks down its ratings to include a ``Tilt'' category. Our SDE model and \cite{538govSen} provide numbers to quantify uncertainty. For toss-ups, the number or text is red (respectively, blue) if it corresponds to a Republican's (respectively, Democrat's) chance of winning. We indicate the actual election outcome by the coloring of the state name. In panel (b), we do not forecast the MN special (MN*) election in July, because this race had no polls before August.
}\label{fig:gov1}
\end{figure}

As we show in Figure~\ref{fig:gov1}, there is a lot of variability in how strongly different sources forecast the races. A helpful measure to evaluate classification models is logarithmic loss \cite{logloss}, which rewards confident forecasts that identify the winning candidate and penalizes confident forecasts that do not identify the winner. It is given by
\begin{align*}
	\text{log loss} &= -\frac{1}{E} \sum_{j=1}^E \left( y_i \log p_i + \left(1-y_i\right) \log \left(1-p_i\right) \right)\,,
\end{align*}
where `log' is the natural logarithm, $E$ is the number of states that we treat individually (so $E = 13$ and $E = 14$ for the 2018 gubernatorial and senatorial races, respectively), $y_i = 1$ if the projected candidate wins in state $i$ and $y_i = 0$ otherwise, and $p_i$ is the probability (see the percentages in Figure~\ref{fig:gov1}) that we assign to the projected winning candidate in state $i$. To calculate the log-loss error for qualitative forecasts, we specify $p_i = 0.5$ for ``Toss-up'', $p_i = 0.55$ for ``Tilt'', $p_i = 0.675$ for ``Lean'', $p_i = 0.85$ for ``Likely'', and $p_i = 0.975$ for ``Solid''. 
As we show in Table \ref{table:success2018}, the forecasts {from our SDE model~\eqref{eq:sde1}--\eqref{eq:sde3}} rank second and third for the races for senator and governor, respectively, among our example popular forecasters according to log-loss error. {In comparison to the log-loss errors in Table \ref{table:success2018}, a log-loss error of $0.7$ corresponds to a hypothetical forecast that assigns a 50\% chance to each of the two candidates in a race.

We have focused predominantly on producing final forecasts (i.e., those that are available right before an election takes place), in part because public attention often centers on how forecasts from the eve of an election compare to race outcomes and in part because our work is a first step toward data-driven election forecasting from a dynamical-systems perspective. However, the most meaningful forecasts are those in the weeks and months before an election day, {and there is particular value in forecasts that remain stable across time \cite{Linzer,Wang,Lewis-Beck}}. {Producing an early forecast is particularly challenging, and it provides a more comprehensive view of a model's worth \cite{Linzer}. To begin to address these ideas, we show earlier forecasts for the 2018 races in Figure~\ref{fig:gov1}. We base these forecasts on less polling data; for example, our $8$~July forecast uses polling data up to and including $8$~July. We use the same superstate categorizations in these forecasts as we do in our final forecasts, which rely on the ratings of popular forecasters in August and November (see Section \ref{super}). Notably, our July, August, September, and October forecasts for swing states in the senatorial races are as accurate at calling race outcomes as our final forecasts (see Figure~\ref{fig:gov1}b). As the election nears and we incorporate more polling data into our model, our performance (as measured by log-loss error and {MOV} error) improves. {This supports observations \cite{Campbell,Wlezien} that polling data becomes more reliable over time.} In comparison to Table~\ref{table:success2018}, we miss the true vote margin by $6.7$, $5.8$, and $5.0$ percentage points on average in August, September, and October, respectively. Similarly, our log-loss error decays in time; it is $0.56$ in August, $0.53$ in September, and $0.44$ in October. For the gubernatorial elections, our SDE model calls the same state outcomes roughly two weeks before the election as it does in November.

%%%%%

\section{Conclusions} \label{conclusions}
We developed a method for forecasting elections by adapting ideas from compartmental modeling and epidemiology; and we illustrated the {utility} of such a dynamical-systems approach by applying it to the U.S. races for president, senator, and governor in 2012, 2016, and 2018. When making our modeling choices, we tried to {limit the number of election-specific details in our methodology.} Despite our approach of using poll data without any weighting adjustments, as well as clear differences between voting dynamics and the spread of infectious diseases, we performed similarly to popular forecasters in calling final race outcomes. {Moreover, we were able to forecast the outcomes of the senatorial elections in 2018 using polling data prior to August 2018 with the same {success rate} as FiveThirtyEight's final forecast \cite{538govSen}}.

We consider our model's generality and {basis in the well-studied, multidisciplinary field of mathematical epidemiology} a virtue in this initial dynamical-systems effort, as part of our goal is to help demystify election prediction, highlight future research directions in the forecasting of elections (and other complex systems), and motivate a broader research community to engage 
actively with pollster interpretations and polling data. There are many ways to build on our basic modeling approach and more realistically account for voter interactions.

To give one example of a viable research direction, it will be useful to be more nuanced about how to handle undecided and minor-party voters. FiveThirtyEight \cite{538methodPres} assigns a voting opinion (mostly to one of the major parties) to any undecided voters who remain on election day, and the \emph{The Huffington Post} factors undecided voters into an election's uncertainty \cite{HuffPostMethod}. {By contrast, we assumed that any undecided voters 
at the end of our simulations are minor-party voters.} Using the fraction of undecided voters to inform one's choice of noise strength is an interesting direction to pursue. {Moreover, because we compare our model with some popular qualitative forecasters, we showed our forecasts as projected vote margins rather than as absolute Republican and Democrat percentages. It may be desirable for future studies to look more closely at how the fractions of Republicans, Democrats, and undecided voters evolve in time (see, e.g., \cite{Linzer,Xbox}). }

We assumed that all polls are equally accurate (e.g., we did not consider the time to election and pollster-reported error), and we did not distinguish between partisan and non-partisan polls or between polls of likely voters, registered voters, and adults. This minimal, poll-aggregating approach echoes the work of Wang \cite{Wang}. By contrast, FiveThirtyEight \cite{538methodPres} relies on measures \cite{538ratings} of polling-firm accuracy to weight polls, and its analysts adjust polls of registered voters and adults to frame all of their data in terms of likely voters. Using FiveThirtyEight's pollster ratings \cite{538ratings} to weight polls in our model would allow us to explore how the various subjective choices of forecasters determine their predictions. Similarly, future work can compare the influence of noise that is correlated based on demographics with the effects of noise that is correlated based on roughly the last $80$ years of state voting history. \emph{The Huffington Post} \cite{HuffPostMethod} uses the latter (and it does not use demographics).

{In our study of election dynamics, we took a macroscopic, simplified view of state and voter interactions. We based our approach on compartmental modeling of contagion spreading because it gives a well-established, multidisciplinary way to include asymmetric state--state relationships in a model. However, when a social behavior or opinion appears to spread in a community, it is often difficult to determine whether transmission is actually occurring. In particular, the appearance of ``spreading" may emerge because social contagions are truly spreading between individuals (that is, individuals are influencing each other), because people form relationships with others who are similar to them and behave in a similar way (e.g., adapting the same opinion) due to their shared characteristics \cite{Homophily}, because of some external factors, or because of a combination of such processes. By building more detailed mathematical models of voter behavior in the future, one can help elucidate what role influence plays in political opinion dynamics. }

Our models assume that every voting-age individual is equally likely to interact with any other voter in the U.S. Although this mean-field approach fits within the theme of simplicity that we embraced throughout our modeling process, the assumption of uniform mixing is not particularly realistic \cite{Meyers, Watts,Busenberg}. For example, several celebrities were heavily involved in encouraging individuals to vote in the 2018 elections, and they have more prominent platforms than a typical voter. Accounting for realistic network structures and exploring frameworks other than compartmental models --- such as voter models \cite{PhysRevLett.112.158701,Braha}, local majority-rule models \cite{GALAM200456,GalamTrump}, and threshold models \cite{LehmannSune2018} --- {may be helpful for capturing relationships between voters. Network models may also allow future studies of how different methods, such as ``big nudging"  \cite{SEME, Harvard}, may influence voter turnout and behavior at the individual level.}

In future modeling efforts, it will also be useful to incorporate additional types of data (e.g., measurements of partisan prejudice by county \cite{GeographyPartisan}), as they become available, into election models to improve both the detail and the quality of forecasts. Our models work on the level of states because state polling data is available; by contrast, House elections are polled less regularly, making fundamental data more important for these races \cite{538methodSen}. Although {precinct-level data} is available on presidential election results {(e.g., see \cite{LATimesData})}, we are not aware of polls at the precinct level. Because many states have regions with different voting behavior, such as urban versus rural areas, precinct-level polling data has the potential to lead to significant improvements in forecasts.

When fitting our parameters, we averaged polls \cite{RealClearPoliticsData,HuffPostPollster} by month (see Section \ref{fit}). This technique smooths out daily fluctuations, {which may be more representative of sampling error than of real shifts in opinion \cite{Jackman}, so it may throw out certain interesting dynamics, such as those that occur around party conventions \cite{Linzer}.} (As described by FiveThirtyEight \cite{538methodPres}, candidates often receive a spike in support after their party's convention.) {The impact of campaigns and media coverage on opinion dynamics is debated in the political-science community \cite{Gelman,Wlezien,Campbell}. With a finer view on polls, one can explore the possible effects of time-specific events, such as a large rally or a story about a candidate in the media, using our modeling framework.} Similarly, one can build feedback mechanisms into a model to test how the perception of future election results influences an individual's likelihood of voting. The framework of dynamical systems provides a valuable approach for exploring the temporal evolution of opinions and their interplay with external forces (such as the media, rallies, and conventions). 

Our compartmental-model approach allowed us to obtain parameters that are related to the strengths of interactions between states and measurements of voter turnover by state for each election year and race (see Figures~\ref{fig:par1}--\ref{fig:par3}). {In the future, it will be useful to investigate the stability of our forecasts and parameters under alternative methods (e.g., data-assimilation methods} \cite{Stuart}) {for fitting these parameters.} By comparing our parameters across years, various types of elections, {and different approaches for fitting}, one can help identify blocs of states that are related persistently, analyze which states have the most plastic voter populations, and suggest differences in the political dynamics in presidential, senatorial, and gubernatorial races. {One can also use our parameters from previous elections to provide early forecasts for upcoming races prior to when large-scale polling data becomes available.} These and other future research directions may provide insight into how state relationships evolve across years, allowing researchers to identify ways that the U.S. electorate may be changing in time, which may in turn suggest ideas to incorporate into future forecasts.

%%%%%

{\appendix{

\section{Additional background on compartmental models} \label{compartmental}
The dynamics of transmission and recovery in a susceptible--infected--susceptible (SIS) model are described by the following coupled ordinary differential equations:
\begin{align*} 
	\frac{d\tilde{S}}{dt} &= \gamma \tilde{I} - \tilde{\beta} [\tilde{S}\tilde{I}]\,, \\
	\frac{d\tilde{I}}{dt} &= - \gamma \tilde{I} + \tilde{\beta} [\tilde{S}\tilde{I}]\,,
\end{align*}
where $\tilde{S}(t)$ and $\tilde{I}(t)$, respectively, are the mean numbers of susceptible and infected individuals in a population at time $t$ and $[\tilde{S}\tilde{I}]$ is the mean number of connections between infected and susceptible individuals. To close the system \cite{kiss2017}, we use the approximation $[\tilde{S}\tilde{I}] \approx \tilde{S} \cdot n \cdot \tilde{I}/N$, where $N$ is the total number of individuals in a population and $n$ is the mean number of connections per person. We define the notation $\beta := \tilde{\beta}n$ and obtain the following system:
\begin{align*} 
	\frac{d\tilde{S}}{dt} &= \gamma \tilde{I} - \beta \tilde{S}\tilde{I}/N \,, \\
	\frac{d\tilde{I}}{dt} &= - \gamma \tilde{I} + \beta \tilde{S}\tilde{I}/N\,.
\end{align*}
Writing this system in terms of the population fractions, $S(t) = \tilde{S}(t)/N$ and $I(t) = \tilde{I}(t)/N$, and dividing by $N$ yields Eqns.~\eqref{eq:SIS1}--\eqref{eq:SIS2}. 

A similar process yields our two-pronged deterministic SIS model in Eqns.~\eqref{eq:elec1}--\eqref{eq:elec3} for election forecasting. The key difference is how we calculate
$[S^iI^j_\text{D}]$ and $[S^iI^j_\text{R}]$. Specifically, we use the approximation
\begin{align*} 
	[\tilde{S}^i\tilde{I}^j_\text{D}] &\approx \tilde{S}^i \cdot n \cdot \left(N^j/N\right) \cdot \left(\tilde{I}^j_\text{D}/N^j \right) \,, \\
	[\tilde{S}^i \tilde{I}^j_\text{R}] &\approx \tilde{S}^i \cdot n \cdot \left(N^j/N\right) \cdot \left(\tilde{I}^j_\text{R}/N^j \right) \,.
\end{align*}
We thereby estimate, for example, the number of interactions between undecided voters in state $i$ and Democrats in state $j$ as the mean number of interactions that involve an undecided voter in state $i$ multiplied by the probability that the interaction is with someone in state $j$ multiplied by the probability that someone in state $j$ is a Democrat. In making these approximations, we are assuming that an undecided voter is equally likely to interact with a Republican or a Democrat across the U.S. In particular, we do not assume that interactions are more likely between individuals in the same state or between those in neighboring states. We also ignore any effects of homophily (even though people are more likely to interact with others who are similar in some way, such as political outlook \cite{newman2018}) and that the number of interactions between individuals in different states depends only on the voting-age populations \cite{FedRegister2016,FedRegister2017,FedRegister2012} of the states and on the number of Republicans, Democrats, and undecided voters currently therein.

%%%
\section{Election-modeling details}
We now provide additional details about our election forecasting process. We overview the data that we use and describe several special cases in Section \ref{data}. We then discuss how we select superstates (Section \ref{super}) and how we numerically implement our model (Section \ref{numerical}).

%%%
\subsection{Data} \label{data}
We obtained publicly-available state polling data for 2012 and 2016 from HuffPost Pollster \cite{HuffPostPollster} using the Pollster API v2 \cite{HuffPostAPI}. State polling data for 2018 was not available from HuffPost Pollster \cite{HuffPostPollster}, so we collected 2018 data by hand from RealClearPolitics \cite{RealClearPoliticsData}. See our GitLab repository \cite{Gitlab_elections} for the 2012, 2016, and 2018 polling data. We use 2012, 2016, and 2017 estimates of voting-age population sizes from the Federal Register \cite{FedRegister2012, FedRegister2016,FedRegister2017} to specify $N$ and $N^i$ in Eqns.~\eqref{eq:elec1}--\eqref{eq:sde3}; we use 2017 data for 2018, because 2018 measurements were not yet available at the time of our analysis.

%%%

\subsubsection{Special cases and notes} \label{special}
Working with election data is often
messy, and we comment on a few special cases in our efforts. 
\begin{itemize}
\item[(1)]{Different election days: Unlike the other 2012 races, the Wisconsin gubernatorial election took place in June 2012, so we do not forecast this race.} 
\item[(2)]{Single-party races: California had two Democrats running for senator in 2018 and 2016. Because our model assumes a race of a Democrat facing a Republican, we do not use polling data from California when it has a single-party race; {naturally, we also do not} forecast these races.}  
\item[(3)]{Independent candidates: The Vermont senatorial races featured an Independent running in place of a Democrat in 2012 and 2018. Consequently, for the 2012 election, we do not forecast Vermont. For the 2018 race, we treat the Independent as a Democrat in our models so that we can still provide a forecast for Vermont. } 
\item[(4)]{Third-candidate polling data: We focus on polling data that compares two candidates. In particular, we do not include polls {that report data for races in which there are three or more candidates who each get reasonably large shares of the vote.} The polls that we found from RealClearPolitics \cite{RealClearPoliticsData} for New Mexico's 2018 senatorial race were for three candidates, so we do not include New Mexico's polls in our averaged data points for the Safe Blue superstate for this election. We also do not forecast the Maine 2012 senatorial race because it included three popular candidates.}
\item[(5)] No polls: In some elections, one or more states have no polls. If these states lie in our Safe Red or Safe Blue superstates, this is not an issue, as we simply assign the vote margin of the appropriate superstate to them. However, in elections for which we forecast each state individually, we cannot forecast states without polling data. Therefore, because polling data from HuffPost \cite{HuffPostPollster,HuffPostAPI} were not available for the 2012 gubernatorial races in Delaware and West Virginia, we do not provide forecasts for these races. 
\item[(6)] {Early forecasts:} {For our $8$ July forecasts for the 2018 senatorial races in Figure~\ref{fig:gov1}b, the Minnesota special election had no polls prior to $8$ July, so we remove this race from our models for this forecast only.}
\item[(7)] {Demographic estimates: To correlate noise in our model} \eqref{eq:sde1}--\eqref{eq:sde3}, {we use estimates of the numbers of Hispanic individuals and non-Hispanic Black individuals in each state in 2016. We gathered these estimates from the U.S. Census Bureau through American FactFinder} \cite{census}. {Later, after the development of our model, American FactFinder was decommissioned, so the website in} \cite{census} {is no longer available. Demographic estimates are now available at} \cite{census2}, {but this data is slightly different than the estimates that we used, because the U.S. Census Bureau revises their past estimates when they make new estimates. See \cite{Gitlab_elections}} {for the data that we used.}
\end{itemize}

%%%%%

\subsection{Selecting superstates} \label{super}
We focus on forecasting elections in swing states and treat all reliably Red and Blue states together as two ``superstate'' conglomerates. (We do not specify that these superstates actually vote Republican and Democrat, respectively; such voting results are outputs of our models.) This raises the question of how to identify states as ``safe'' or ``swing'', and we do this differently for different elections. For presidential races, we define our swing states as the ones that FiveThirtyEight has identified as ``traditional swing states'' \cite{swing}; these are Colorado (CO), Florida (FL), Iowa (IA), Michigan (MI), Minnesota (MN), Nevada (NV), New Hampshire (NH), North Carolina (NC), Ohio (OH), Pennsylvania (PA), Virginia (VA), and Wisconsin (WI) (see Figure~\ref{fig:overview}d). Therefore, for the presidential elections, $M=14$ in Eqns.~\eqref{eq:elec1}--\eqref{eq:sde3}. In our notation, $\{S^1,I^1_\text{D},I^1_\text{R}\}$ and $\{S^2,I^2_\text{D},I^2_\text{R}\}$ refer to the voter fractions in the Red and Blue superstates, respectively; and $\{S^i,I^i_\text{D},I^i_\text{R}\}$ for $i \in \{3,4,\ldots,14\}$ are the voter fractions in the $12$ swing states.

To define superstates in the races {for senator and governor}, we use the race ratings of popular forecasters. For the 2018 senatorial races, we combine the August 2018 ratings of Sabato's Crystal Ball \cite{sabatoSenate}, 270toWin (the consensus version) \cite{270toWin}, and the \emph{New York Times} \cite{NYTsenate}. We determine which states to include in our superstates for the 2018 gubernatorial races based on the ratings of FiveThirtyEight \cite{538govSen}, the Cook Political Report \cite{cookGov}, Sabato's Crystal Ball \cite{sabatoGov}, and Inside Elections \cite{insideSenGov} (all accessed on 1~November 2018). We define our Safe Red and Safe Blue superstates for the 2012 senatorial races based on Sabato's Crystal Ball \cite{sabato2012} and the \emph{New York Times} \cite{NYTsen2012}. We base our superstates for the 2016 senatorial races on 270toWin \cite{270toWin}, Sabato's Crystal Ball \cite{Sabato}, and \emph{The Huffington Post} \cite{HuffPostMethod}. We treat each state separately for the 2012 and 2016 gubernatorial races. In Table~\ref{table:s1}, we give a summary (for each election) of the states that we forecast individually and those that we combine into the Safe Red and Safe Blue superstates.

%%%%

\subsection{Numerical implementation}  \label{numerical}
For our parameter fitting, we use the {\sc optim} routine in R (version 3.4.2) \cite{R} to perform constrained optimization of the least-squares objective function, subject to non-negative rate constraints \cite{byrd1995limited}, with a time step of $\Delta t = 3$ days over $T$ months. (As a simplification, we assume that each month is $30$ days long.) We use this time step in all cases, except for the forecasts in Figures~\ref{fig:senateSI2}b,c (for which we use $\Delta t = 15$ days). We simulate our models in {\sc Matlab} (version 9.3). For each state (or superstate), we set its initial condition to the earliest of its $T$ data points that we use for parameter fitting. We solve our ODE model~\eqref{eq:elec1}--\eqref{eq:elec3} using a forward Euler scheme and our SDE model~\eqref{eq:sde1}--\eqref{eq:sde3} using the Euler--Maruyama method \cite{Higham}. We use the constraint that $S^i+I^i_\text{R}+I^i_\text{D}=1$ to reduce our system to $2$ equations per state (or superstate) for our simulations. For both of our models, we use a time step of $\Delta t = 0.1$ day, and we simulate them from 1~January until an election day. In these simulations, we assume that each month has a length of $30$ days. Specifically, we simulate for $306$ days for the 2012 and 2018 races and for $308$ days for the 2016 races. The noise strength in our SDE system~\eqref{eq:sde1}--\eqref{eq:sde3} is $\sigma = 0.0015$ in all of our simulations.

%%%%
}}

\FloatBarrier

%%%%%
\newpage

\section*{Supplementary Materials}

\section{The original forecasts that we posted on 5~November 2018} \label{original}

We posted our forecasts for the 2018 senatorial and gubernatorial races on the arXiv preprint server at \cite{preprint} on 5~November, the eve of the midterm elections that were held on 6~November 2018. We collect these original forecasts in Figures~\ref{fig:senateSI2}--\ref{fig:govSI2}. After checking our results without the election-time rush, we found that we made some errors when gathering the data. These errors, which we have corrected in the forecasts that we present in the main manuscript\footnote{Note, however, that our original forecasts and those in the main manuscript forecast the same candidates to win each race.}, are as follows:
\begin{itemize}
\item We incorrectly copied the results for the last Rhode Island senatorial poll, which took place 20--24 October 2018. Because we include Rhode Island in our ``Safe Blue'' superstate, this error led to a difference of less than $0.5$ percentage points in the Republican and Democrat percentages at one time point in the ``Safe Blue'' data that we used to fit model parameters.
\item We neglected to incorporate the only Washington poll in our senatorial data into the polling data that we averaged for the ``Safe Blue'' superstate.
\item We incorrectly copied the Republican vote share for the last Maine gubernatorial poll, which took place 27--29 October; instead of using 42\%, we used 37\%.
\item {Instead of correlating noise for each state in our SDE model~\eqref{eq:sde1}--\eqref{eq:sde3} using its associated demographic information as described in Section \ref{sec:trump}, we incorrectly ordered the demographic information in one file, such that we were not associating the correct demographic data to each state. This error does not have a strong impact on our forecasts, suggesting that the critical point is to correlate noise in some way and that this way does not necessarily need to be by demographics.}
\end{itemize}
{These errors occur only in Figures~\ref{fig:senateSI2}, \ref{fig:govSI1}, and \ref{fig:govSI2}.} Additionally, because of time constraints, we generated our {final} senatorial forecasts (based on data from \cite{RealClearPoliticsData} that we gathered through 3~November), which we show in Figures~\ref{fig:senateSI2}b and \ref{fig:senateSI2}c, {using a time step of $0.5$ months for parameter fitting and based on only $4,000$ simulated elections. In all other cases, we specify a time step of $0.1$ months (i.e., $3$ days, because we assume that all months are $30$ days long) for parameter fitting and base our forecasts on $10,000$ simulated elections. Note that we use our deterministic model~\eqref{eq:elec1}--\eqref{eq:elec3} to fit parameters and our stochastic model~\eqref{eq:sde1}--\eqref{eq:sde3} to simulate the 2018 elections.}

 \begin{figure}[t]
 \centering
 \includegraphics[width=1\textwidth]{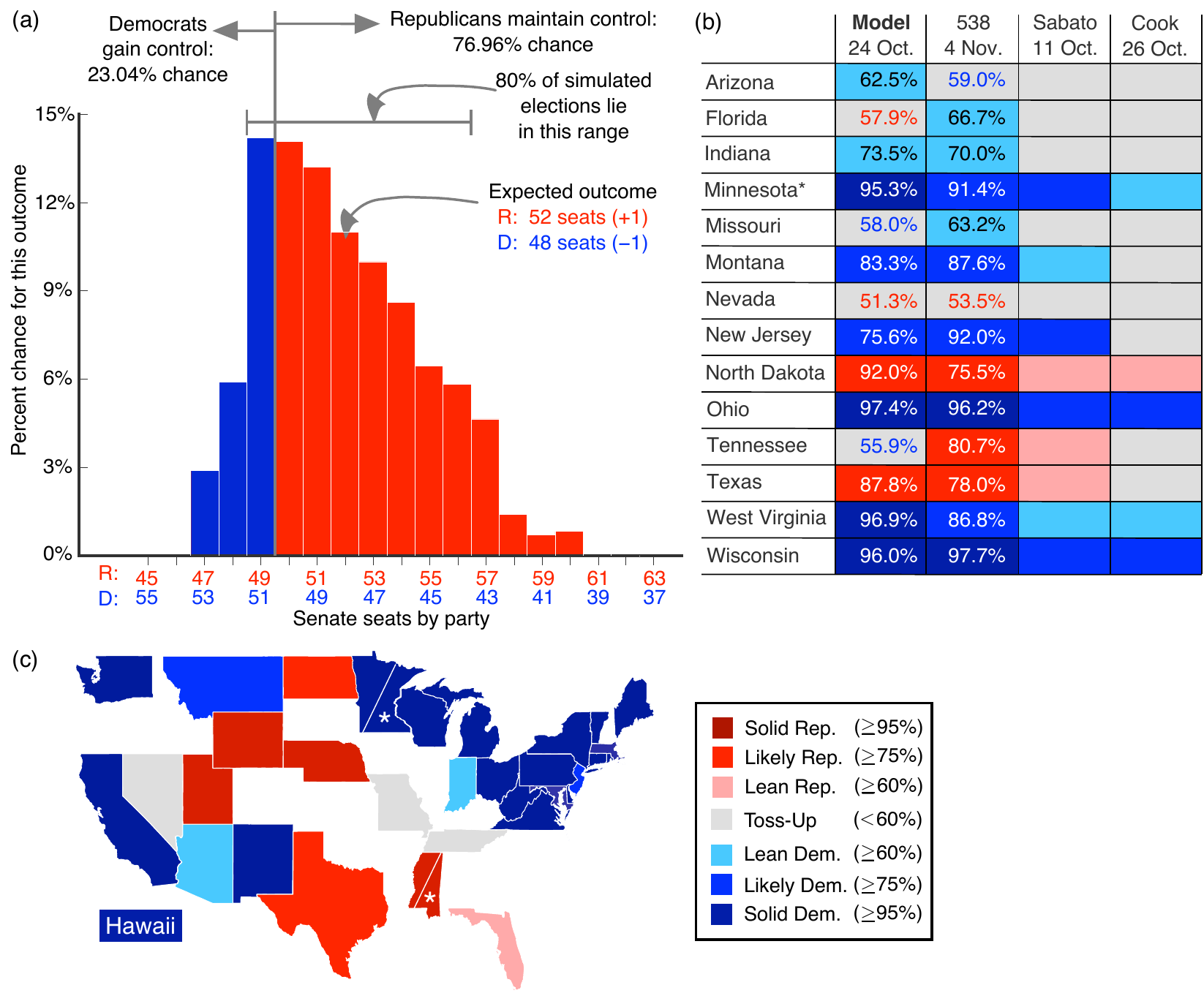}
 \caption{{Our original 2018 senatorial forecasts, which we posted on 5 November 2018 \cite{preprint}.}  (a) Forecast for the composition of the Senate based on our stochastic model~\eqref{eq:sde1}--\eqref{eq:sde3} (on 24~October 2018) using correlated noise. (b) State ratings as Solid, Likely, Lean, or Toss-up from our stochastic model~\eqref{eq:sde1}--\eqref{eq:sde3}, FiveThirtyEight \cite{538govSen}, Sabato's Crystal Ball \cite{sabatoSenate}, and the Cook Political Report \cite{cookGov}. We use ``*'' to designate the Minnesota special election. We generated our 3 November forecasts using polls from \cite{RealClearPoliticsData} through 3~November 2018. (We only rate states that we have not already designated as ``Safe Red'' or ``Safe Blue''; see Table \ref{table:s1} for details.) Each number indicates the chance of a Democrat (respectively, Republican) winning if it is in a blue (respectively, red) box. For toss-up states, each number is red (respectively, blue) if it corresponds to a Republican's (respectively, Democrat's) chance of winning. FiveThirtyEight's values are based on the ``classic'' version of its model that was updated at 1:19 PM Eastern time on 4~November. (c) Map of the state ratings as forecast by our stochastic model using data through 3~November. {We generate our forecasts by simulating our stochastic model~\eqref{eq:sde1}--\eqref{eq:sde3} $10,000$ times for panel (a) and the first column of panel (b) and $4,000$ times for panel (c) and the last column of panel (b).} {For the forecasts in panel (c) and the rightmost column of panel (b), we use a time step for parameter fitting that is five times larger than what we use in our other forecasts.}   
 }
 \label{fig:senateSI2}
 \end{figure}

\begin{figure}[t]
\centering
\includegraphics[width=\textwidth]{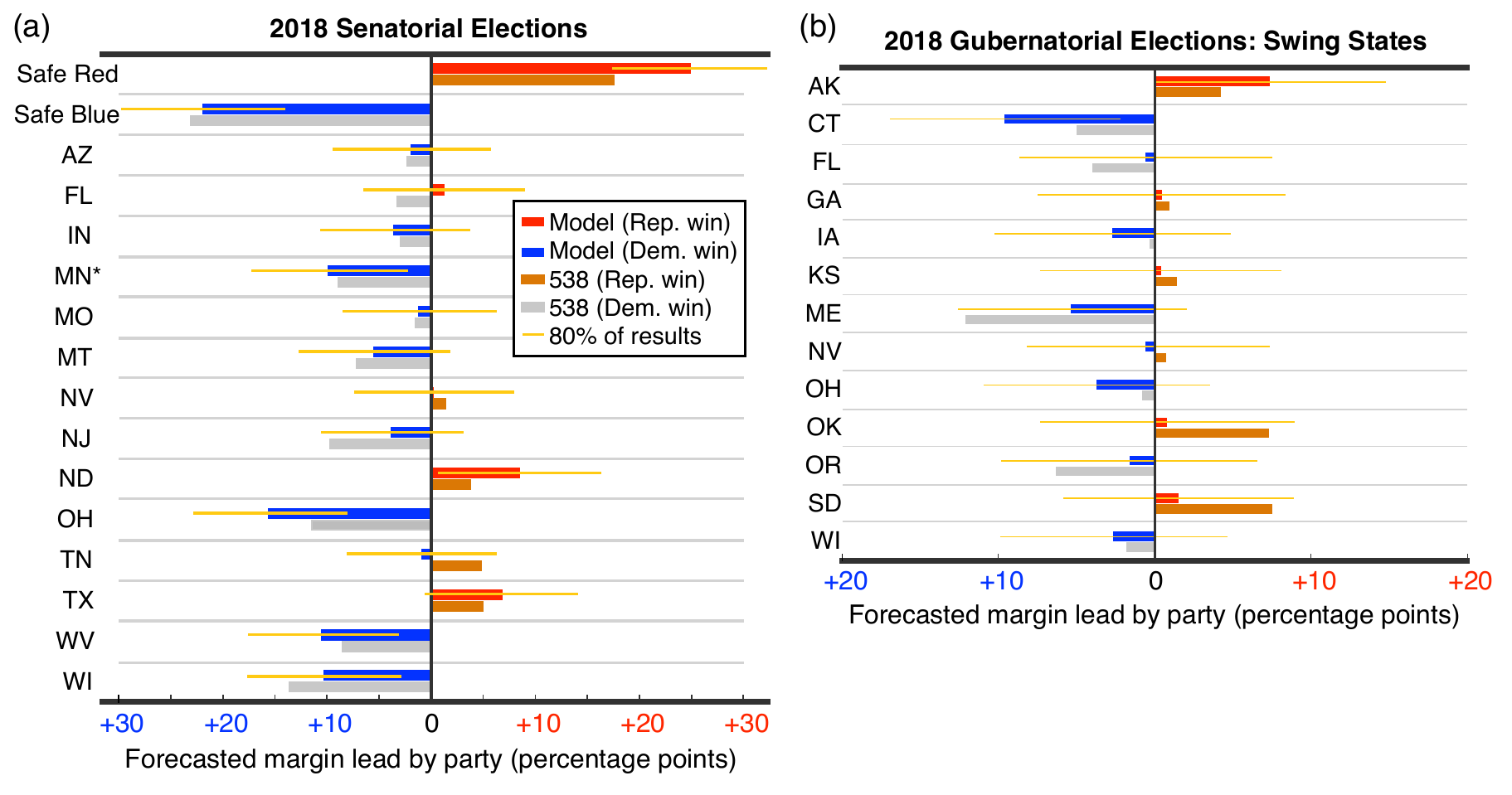}
\caption{{Vote margins for our original 2018 forecasts that we posted on 5 November 2018 
\cite{preprint}.} (a) We show our forecasts (which we based on data that we collected from RealClearPolitics \cite{RealClearPoliticsData} through 24 October 2018) versus those of FiveThirtyEight \cite{538govSen} for the 2018 senatorial races. (We show forecasts from the ``classic'' FiveThirtyEight model; we obtained them on 30~October at 1:06 PM Eastern time.) We use ``*'' to designate the Minnesota special election. See Table \ref{table:s1} for a summary of the states that we include in our Safe Red and Safe Blue superstates. (b) We base our 2018 gubernatorial forecasts on polling data that we collected from RealClearPolitics \cite{RealClearPoliticsData} through 3~November. We also show the forecasts of FiveThirtyEight \cite{538govSen} on 4~November 2018. (The margins that we reproduce from FiveThirtyEight's website are from the ``classic'' version that was updated at 4:50 PM Eastern time on 4 November.) We show the expected margins only for swing states; we forecast the mean margin in the Safe Red superstate to be $+18.5$ points for Republicans and the mean margin in the Safe Blue superstate to be $+15.6$ points for Democrats. {We generate our forecasts in panels (a) and (b) by simulating $10,000$ realizations of our stochastic model~\eqref{eq:sde1}--\eqref{eq:sde3}. The bars signify the mean vote percentages across these stochastic simulations.}\label{fig:govSI1}}
\end{figure}

\begin{figure}[t]
\centering
\includegraphics[width=\textwidth]{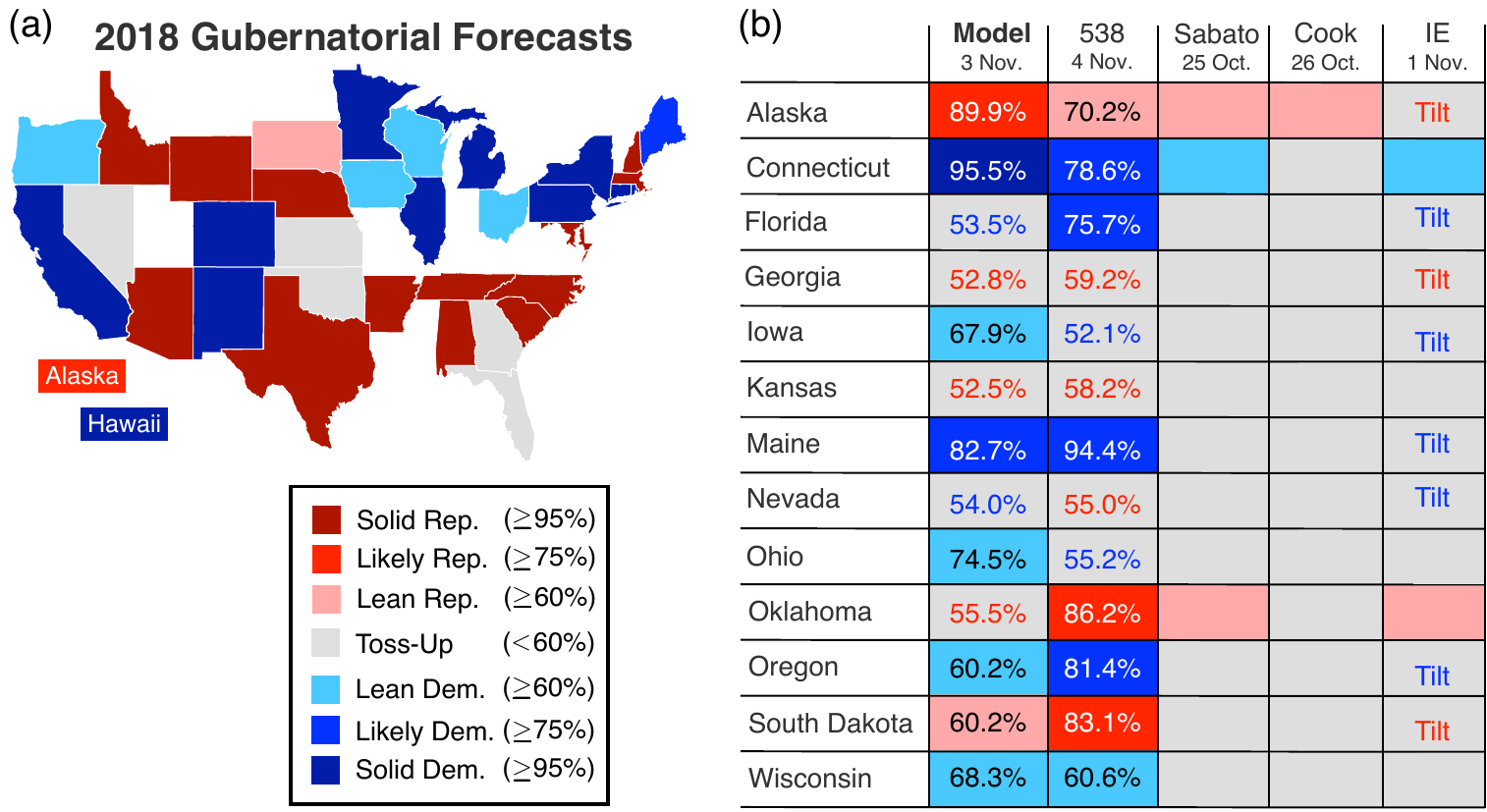}
\caption{{Our original 2018 gubernatorial forecasts that we posted on 5 November \cite{preprint}.} {We generate our forecasts by simulating $10,000$ realizations of our stochastic model~\eqref{eq:sde1}--\eqref{eq:sde3}. (a) Illustration of our gubernatorial forecasts, which we based on data from RealClearPolitics \cite{RealClearPoliticsData} through 3~November 2018. (On the map, we show only states that held gubernatorial elections.) (b) We compare our ratings of states as Solid, Likely, Lean, or Toss-up with those of popular forecasters. (We rate only states that we have not already assigned to the Safe Red or Safe Blue superstates; see the caption of Figure~\ref{fig:govSI1}b for details.) The forecasts that we reproduce from FiveThirtyEight \cite{538govSen} are for the``classic'' version, which was updated at 1:19 PM Eastern time on 4~November 2018. We obtained the forecasts that we reproduce from Sabato's Crystal Ball \cite{sabatoGov}, the Cook Political Report \cite{cookGov}, and Nathan Gonzales's Inside Elections (IE) \cite{insideSenGov} from their websites on 4~November. They were last updated online on 26 October, 26~October, and 1~November, respectively. IE \cite{insideSenGov} further breaks down its state ratings to include a ``Tilt'' category, which lies between the ``Toss-Up'' and ``Lean'' ratings. }
} \label{fig:govSI2}
\end{figure}

%%%%%

\section{Alternative ways of measuring forecast accuracy} \label{alternative}

Because it is not straightforward to measure forecast accuracy \cite{prosser_mellon_2018}, it is important to discuss how subjective choices enter the forecasting process and influence how we view different forecasts. In this section, we highlight a few alternative choices (which are related to Tables~\ref{table:success} and \ref{table:success2018} in the main manuscript) that we could have made when evaluating forecast accuracy.

\begin{landscape}
\begin{table}[t]\centering
\caption{Summary of how we account for each state race by election and year. Depending on the election, we combine some states into Safe Red and Safe Blue superstates (see Section \ref{super}). In some cases, we do not provide a forecast for a given state. (For example, we do not forecast state races in which the two main candidates are from the same party.) In all other cases, we forecast each state race individually. {We designate the special elections for senators for Minnesota and Mississippi using asterisks. {For the special case of our July forecasts of the 2018 senatorial races in Figure\ \ref{fig:gov1}c, we forecast only $13$ states individually. (We exclude MN*, although we forecast this election for later dates, because it has no polling data prior to $8$ July.) We use the \# symbol to signify counts in the columns.
 }} \label{table:s1}}
\begin{small}
\begin{tabular}{p{1.2cm} p{5.1cm} p{5.2cm} p{5.1cm} p{2.3cm}}
Election & Forecast individually (\#) & Safe Red superstate (\#) & Safe Blue superstate (\#) & Not forecast (\#) \\
\midrule
2012 {Gub}. & IN, MO, MT, NH, NC, ND, UT, VT, WA (9) & (0) &  (0) & DE, WV, WI (3) \\ \midrule
2012 Sen. & AZ, CT, FL, IN, MA, MO, MT, NV, ND, OH, PA, VA, WI (13) & MS, NE, TN, TX, UT, WY (6) & CA, DE, HI, MD, MI, MN, NJ, NM, NY, RI, WA, WV (12) & ME, VT (2) \\ \midrule
2012 Pres. & CO, FL, IA, MI, MN, NV, NH, NC, OH, PA, VA, WI (12) & AL, AK, AZ, AR, GA, ID, IN, KS, KY, LA, MS, MO, MT, NE, ND, OK, SC, SD, TN, TX, UT, WV, WY (23) & CA, CT, DE, DC, HI, IL, ME, MD, MA, NJ, NM, NY, OR, RI, VT, WA (16) & (0)\\ \midrule
2016 {Gub}. & DE, IN, MO, MT, NH, NC, ND, OR, UT, VT, WA, WV (12) & (0) & (0) & (0) \\ \midrule
2016 Sen. & AZ, FL, IL, IN, LA, MO, NV, NH, NC, OH, PA, WI (12) & AL, AK, AR, GA, ID, IA, KS, KY, ND, OK, SC, SD, UT (13) & CO, CT, HI, MD, NY, OR, VT, WA (8) & CA (1)  \\ \midrule
2016 Pres. & CO, FL, IA, MI, MN, NV, NH, NC, OH, PA, VA, WI (12) & AL, AK, AZ, AR, GA, ID, IN, KS, KY, LA, MS, MO, MT, NE, ND, OK, SC, SD, TN, TX, UT, WV, WY (23) & CA, CT, DE, DC, HI, IL, ME, MD, MA, NJ, NM, NY, OR, RI, VT, WA (16) & (0)\\ \midrule
2018 {Gub}. & AK, CT, FL, GA, IA, KS, ME, NV, OH, OK, OR, SD, WI (13) & AL, AZ, AR, ID, MD, MA, NE, NH, SC, TN, TX, VT, WY (13) & CA, CO, HI, IL, MI, MN, NM, NY, PA, RI (10) & (0) \\ \midrule
2018 Sen. & AZ, FL, IN, MN*, MO, MT, NV, NJ, ND, OH, TN, TX, WV, WI (14) & MS, MS*, NV, UT, WY (5) & CT, DE, HI, ME, MD, MA, MI, MN, NM, NY, PA, RI, VT, VA, WA (15)& CA (1) \\ 
\bottomrule
\end{tabular}
\end{small}
\end{table}
\end{landscape}

{As we note in Table~\ref{table:s1}, we do not forecast single-party races (namely, the races for senator for California in 2016 and 2018). Depending on the forecast goal, such single-party races either lead immediately to forecast success or are as difficult to forecast as any other race. For example, because two Democrats ran in the California senatorial race in 2018, we could have chosen to forecast this as a Safe Blue state. If our goal is to call state results by color, we interpret such a forecast as a success. However, if our goal is to determine the winning candidate in each state, such a forecast is meaningless. When measuring accuracy, this leaves us with several choices. Should we calculate (and potentially inflate) success rates by counting single-party state races as immediate forecast successes? Should we count these single-party races as failures unless a forecaster identifies the winning candidate? Should we calculate forecast accuracy based only on races that have candidates from two or more parties? We employ the third option --- and we thus leave single-party state races out of forecast-accuracy measurements both for our models and for the popular forecasters --- throughout our work. This choice slightly influences the success rates that we report for the 2016 and 2018 senatorial races because California featured a single-party race in both of these years.}

{As we discussed in Section \ref{special}, in a few cases, we do not provide a forecast for a specific state race. These races either have no polling data, included more than two popular candidates, or included an independent candidate who ran against a Republican. (We exclude races that feature an independent candidate versus a Republican in 2012, but we treat the independent candidate as a Democrat in 2018 so that we can still provide forecasts.) These observations also leave us with choices to make when we measure accuracy. There are two options:
 \begin{itemize}[noitemsep,nolistsep]
 \item Option (a): Evaluate our forecasts and those of popular sources by calculating success rates only for the state races that we forecast.
 \item Option (b): Evaluate success rates across all of the (non-single-party) state races (and thus count our model as failing to make a correct prediction for states that we do not even attempt to forecast).
 \end{itemize}
 We present success rates based on option (a) (i.e., using Eqn.~} \ref{eq:success}) in Table~\ref{table:success} in the main manuscript. Under option (a), across the states for which forecasts are available from our model, we correctly forecast $89.1\%$ of $64$ state races for senator in 2012 and 2016, whereas Sabato achieved a success rate of $93.8\%$. Our forecasts have a success rate of $95.2\%$ for the $21$ gubernatorial races that we forecast in 2012 and 2016; for these races, Sabato's forecasts have a $81.0\%$ success rate. If we instead choose option (b), one obtains different success rates for the 2012 gubernatorial and senatorial races. Specifically, because Sabato's Crystal Ball \cite{sabato2012} correctly forecast the Delaware and West Virginia gubernatorial races that we could not forecast due to the lack of polling data, Sabato's 2012 success rate at calling state outcomes for the races for governor increases to 81.8\% (with $9$ state outcomes called correctly out of $11$ races). Note that we leave Wisconsin out of these calculations, because its election was held several months earlier than the other elections. With this approach to measuring accuracy, our 2012 gubernatorial success rate goes down to 81.8\% (with $9$ state outcomes called correctly and $2$ races not forecast). Similarly, because Sabato's Crystal Ball \cite{sabato2012} correctly forecast the 2012 Maine and Vermont senatorial races that we did not forecast, Sabato's success rate increases to 93.9\% (with $31$ state outcomes called correctly out of $33$), and our success rate decreases to 84.8\% (with $28$ state outcomes called correctly and $2$ races not forecast) when we choose to evaluate performance across all state races. Because our current methodology limits us to forecasting races for which polling data are available, measuring success across all states highlights the value of fundamental data, which is used by Sabato's Crystal Ball \cite{sabato2012} to provide forecasts even when there is not polling data.

{Because of our use of superstates, we also need to make subjective choices in our measurements of MOV error in Table~\ref{table:success2018} in the main manuscript. Specifically, when we combine states into Safe Red and Safe Blue superstates, we forecast the mean Democrat and Republican vote percentages across the selected states. This leaves us with three choices for calculating MOV error and comparing it with that of FiveThirtyEight \cite{538govSen}. Option (a) is to apply our strategy to FiveThirtyEight's forecasts. For example, we can calculate a weighted average of FiveThirtyEight's vote margins across the state races in our superstates with weights given by the number of voting-age individuals in each state. This is the technique that we used in Figure~\ref{fig:senateFinal}. Option (b) is to treat the Safe Red and Safe Blue vote margins that we forecast as if they are the forecast margins of each individual state. For example, if we forecast the Safe Red superstate to go $+18$ points Republican, we interpret this result as signifying that each of the individual states in the Safe Red superstate is $+18$ points Republican. Option (c) is to evaluate the MOV error for FiveThirtyEight \cite{538govSen} and our stochastic model~\eqref{eq:sde1}--\eqref{eq:sde3} only on the set of states that we treat individually. We use this third option in Table~\ref{table:success2018}. As alternative measurements, we include MOV errors that we calculate using options (a) and (b) in Table~\ref{table:STable21}. Although all of the states in our Safe Red (respectively, Safe Blue) superstate voted Republican (respectively, Democrat) in the elections that we considered, the margins of victory were rather different across some of these states, even though they all voted for the same party. Therefore, in comparison to FiveThirtyEight \cite{538govSen}, our stochastic model has its worst forecasting performance if we choose option (b), because FiveThirtyEight forecasts each state individually.}

\begin{table*}[t!]
\centering
\caption{Alternative ways of computing the mean error in margin of victory (MOV error) for the 2018 gubernatorial and senatorial races. We show the mean {MOV errors} for our stochastic model~\eqref{eq:sde1}--\eqref{eq:sde3} and FiveThirtyEight \cite{538govSen}. Because of our use of superstates, we identify three ways of measuring MOV error. In option (a), we calculate a weighted average of the vote margins that FiveThirtyEight forecast across the states in our superstates (essentially, we construct FiveThirtyEight's superstate forecasts using its individual state forecasts), and we compute the mean MOV error based on these superstate vote margins and the remaining individual state vote margins. In option (b), we treat the superstate vote margins that we forecast as if they apply to each state individually. For example, if we forecast the Safe Red superstate to go $+18$ points Republican, we interpret this number as signifying that each state in our Safe Red superstate is $+18$ points Republican. This option allows us to compute FiveThirtyEight's mean MOV error based on all of the state races individually. We show the mean MOV error that we compute using option (c) in Table~\ref{table:success2018} in the main manuscript. For option (c), we report the mean MOV error for our stochastic model and FiveThirtyEight based only on the states that we forecast individually. (We do not include the single-party race for senator for California in any of these measurements.) We face analogous options for computing log-loss error. Because these options affect the MOV error and log-less error differently, it is insightful to compare the present table to Table~\ref{table:STable22}. 
 \label{table:STable21}}
\begin{tabular}{lcc}
Forecaster & Our SDE model & FiveThirtyEight \cite{538govSen} \\ \midrule
{Gub}.~error by option (a) & 3.8 pts. & 3.0 pts. \\
{Gub}.~error by option (b) & 5.5 pts. &  3.5 pts. \\
Sen.~error by option (a) & 4.3 pts.  &  3.7 pts. \\
Sen.~error by option (b) & 6.5 pts. & 4.1 pts. \\
\bottomrule
\end{tabular}
\end{table*}

{Because of our use of superstates, we face similar subjective choices for how to calculate log-loss error in Table~\ref{table:success2018} in the main manuscript. We do not specify that the Safe Red (respectively, Safe Blue) superstate be rated as ``Solid Republican'' (respectively, ``Solid Democrat''). Instead, this result is an output of simulating $10,000$ elections with our stochastic model~\eqref{eq:sde1}--\eqref{eq:sde3}. For example, for the 2018 senatorial races, we obtain a Republican outcome for the Safe Red superstate in 100\% of our simulations and a Democrat outcome for the Safe Blue superstate in 100\% of our simulated elections. For the 2018 gubernatorial races, a Republican wins in the Safe Red superstate in 99.86\% of our simulations, and a Democrat wins in the Safe Blue superstate in 99.33\% of our simulations. We can calculate the log-loss error of our forecasts and those of popular sources in a few different ways. These choices are analogous to the options that we discussed above for MOV error. Option (a) is to apply our strategy to the forecasts of popular sources by taking the mean of their individual state ratings to obtain ratings in terms of our superstates. We do not weight these ratings by voting-age population size. Option (b) is to treat our Safe Red and Safe Blue ratings as if they apply individually to each state in the superstates. For example, because 99.86\% of our $10,000$ simulations for the Safe Red superstate have a Republican outcome, we interpret this result as signifying that each individual state in the Safe Red superstate has a 99.86\% chance of voting Republican. Option (c) is to evaluate the log-loss error for our model and popular forecasters only for the set of states that we treat individually. We use option (c) in Table~\ref{table:success2018} in the main manuscript. As alternative measurements, we include MOV errors that we calculate using options (a) and (b) in Table~\ref{table:STable22}.}

\begin{table*}[t!]
\centering
\caption{Alternative ways of computing log-loss error for the 2018 gubernatorial and senatorial races. As we discussed in Section \ref{alternative}, our use of superstates gives several options for computing log-loss error. Briefly, in option (a), we take the mean of the individual state ratings of popular forecasters to obtain their ratings for our Safe Red and Safe Blue superstates. (In this process, we do not weight state ratings by population size.) We then use these superstate ratings, together with the ratings of the states that we treat individually, to compute log-loss error. In option (b), we treat our superstate forecasts as if they apply individually to each state in these superstates. (For example, if the Safe Red superstate goes Republican in 100\% of our simulations of our stochastic model~\eqref{eq:sde1}--\eqref{eq:sde3}, we interpret this result as signifying that each individual state in the Safe Red superstate has a 100\% chance of a Republican outcome.) We showed the log-loss error that we compute using option (c) in Table~\ref{table:success2018} in the main manuscript. For option (c), we reported the log-loss error for our model and popular forecasters only for the states that we forecast individually. We do not include the single-party race for senator for California in the computations of log-loss error for any of these options. We face similar options when computing MOV error; see Table~\ref{table:STable21} for these measurements. } \label{table:STable22}
\begin{tabular}{lc p{2.5cm} p{2.13cm}}
Forecaster & Our SDE model & FiveThirtyEight \cite{538govSen} & Sabato \cite{sabatoSenate, sabatoGov}  \\ \midrule
{Gub}.\ log loss by option (a) & 0.511  & 0.480 & 0.522  \\
{Gub}.\ log loss by option (b) &  0.215 & 0.221 & 0.287 \\
Sen.\ log loss by option (a) &  0.346 & 0.363 & 0.337 \\
  Sen.\ log loss by option (b) & 0.163 & 0.184 & 0.175 \\
\bottomrule
\end{tabular}
\end{table*}

%%%%%%%

\section{Code, data, and model parameters} \label{parameters}

As we discussed in Section \ref{fit}, we determine the model parameters that we use in our deterministic model~\eqref{eq:elec1}--\eqref{eq:elec3} and our stochastic model~\eqref{eq:sde1}--\eqref{eq:sde3} from the main manuscript by fitting Eqns.~\eqref{eq:elec1}--\eqref{eq:elec3} to public polling data \cite{HuffPostPollster,RealClearPoliticsData}. As illustrative examples, we show our transmission and recovery parameters for the 2018 senatorial races in Figures~\ref{fig:par1}--\ref{fig:par3}. We give all of our parameter values in our GitLab repository \cite{Gitlab_elections}. We also provide our model code with detailed instructions on how to reproduce our model parameters \cite{Gitlab_elections}. As we noted in Section \ref{data} in the main manuscript, we base our 2012 and 2016 model parameters on publicly-available polling data from HuffPost Pollster \cite{HuffPostPollster}. In \cite{Gitlab_elections}, we provide this polling data, which we have formatted for use in our models, and the 2018 polling data that we collected by hand from RealClearPolitics \cite{RealClearPoliticsData}. To simulate our stochastic model~\eqref{eq:sde1}--\eqref{eq:sde3}, we use demographic data for the fractions of non-Hispanic Black, Hispanic, and college-educated individuals in each state from the U.S. Census Bureau \cite{census} and \url{247WallSt.com} \cite{education}. We include the demographic data that we have formatted for use in our models from these sources \cite{census,education} in \cite{Gitlab_elections}. Voting-age population sizes $N^i$ (for $i \in \{1,\ldots,M\}$) for each state $i$ are available from the Federal Register \cite{FedRegister2012,FedRegister2017,FedRegister2016} for each election.

\begin{figure}[t]
\centering
\includegraphics[width=\textwidth]{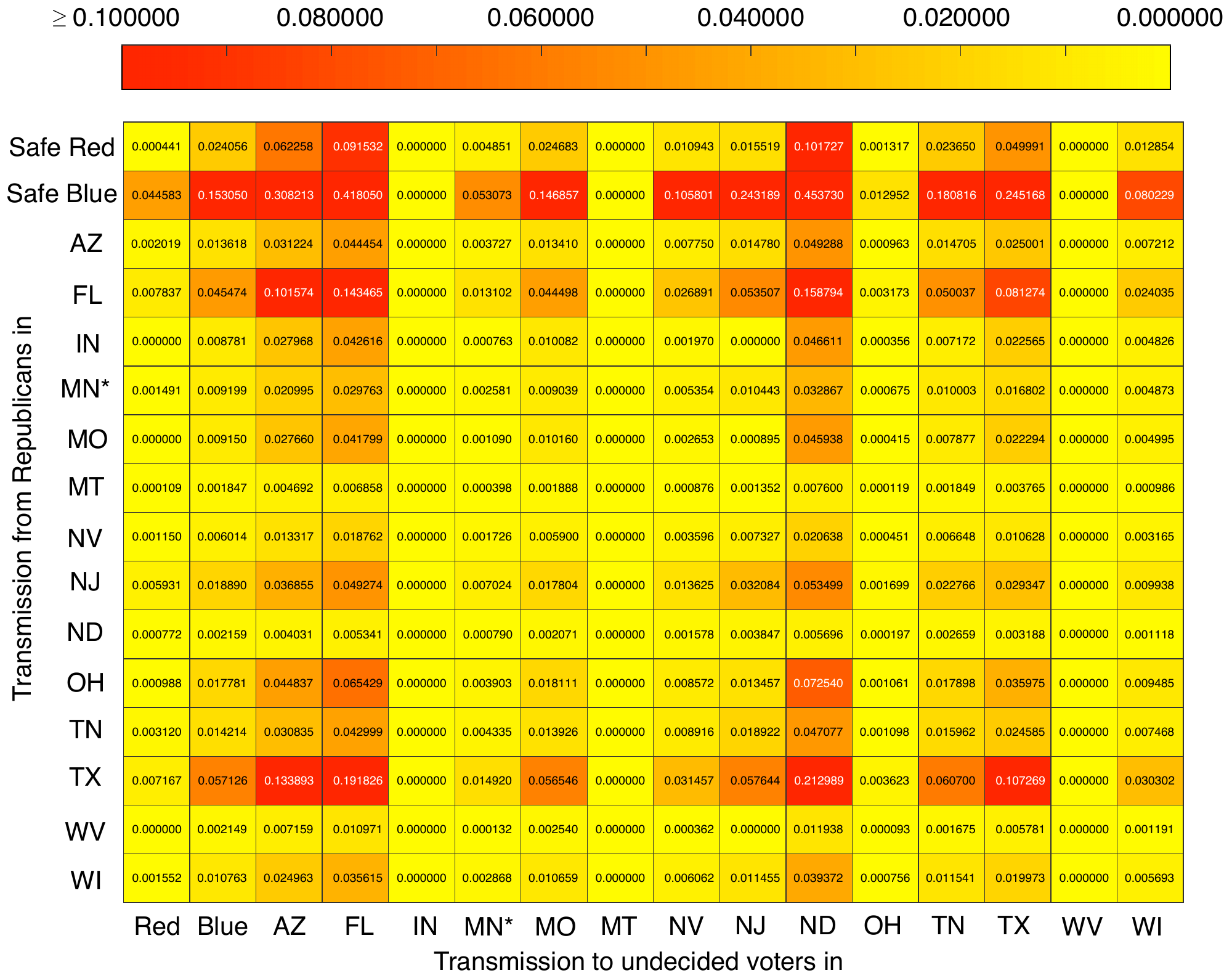}
\caption{{Republican transmission parameters for our 2018 senatorial forecasts.} We provide the parameters {that we use in our stochastic model~\eqref{eq:sde1}--\eqref{eq:sde3} to simulate} elections for our final forecasts in our GitLab repository \cite{Gitlab_elections}. {(In this figure, we show the parameters to $6$ decimal places; see  \cite{Gitlab_elections} for more precise numbers.) } As an example, this figure shows the Republican transmission parameters {(in units of 1/month)} for our 2018 senatorial forecasts using polling data that we obtained from \cite{RealClearPoliticsData} through 3~November 2018. {We calculate these parameters by fitting our deterministic model~\eqref{eq:elec1}--\eqref{eq:elec3} to polling data. To simulate elections, we use these parameters in our stochastic model~\eqref{eq:sde1}--\eqref{eq:sde3}.} The parameter $\beta_\text{R}^{ij}$ describes opinion transmission from Republicans in state $j$ to undecided voters in state $i$. To clarify our notation, we highlight $\beta_\text{R}^\text{FL,TX}$ in the table; this parameter describes the influence that Republicans in Texas have on undecided voters in Florida. \label{fig:par1}
}
\end{figure}

\begin{figure}[t]
\centering
\includegraphics[width=\textwidth]{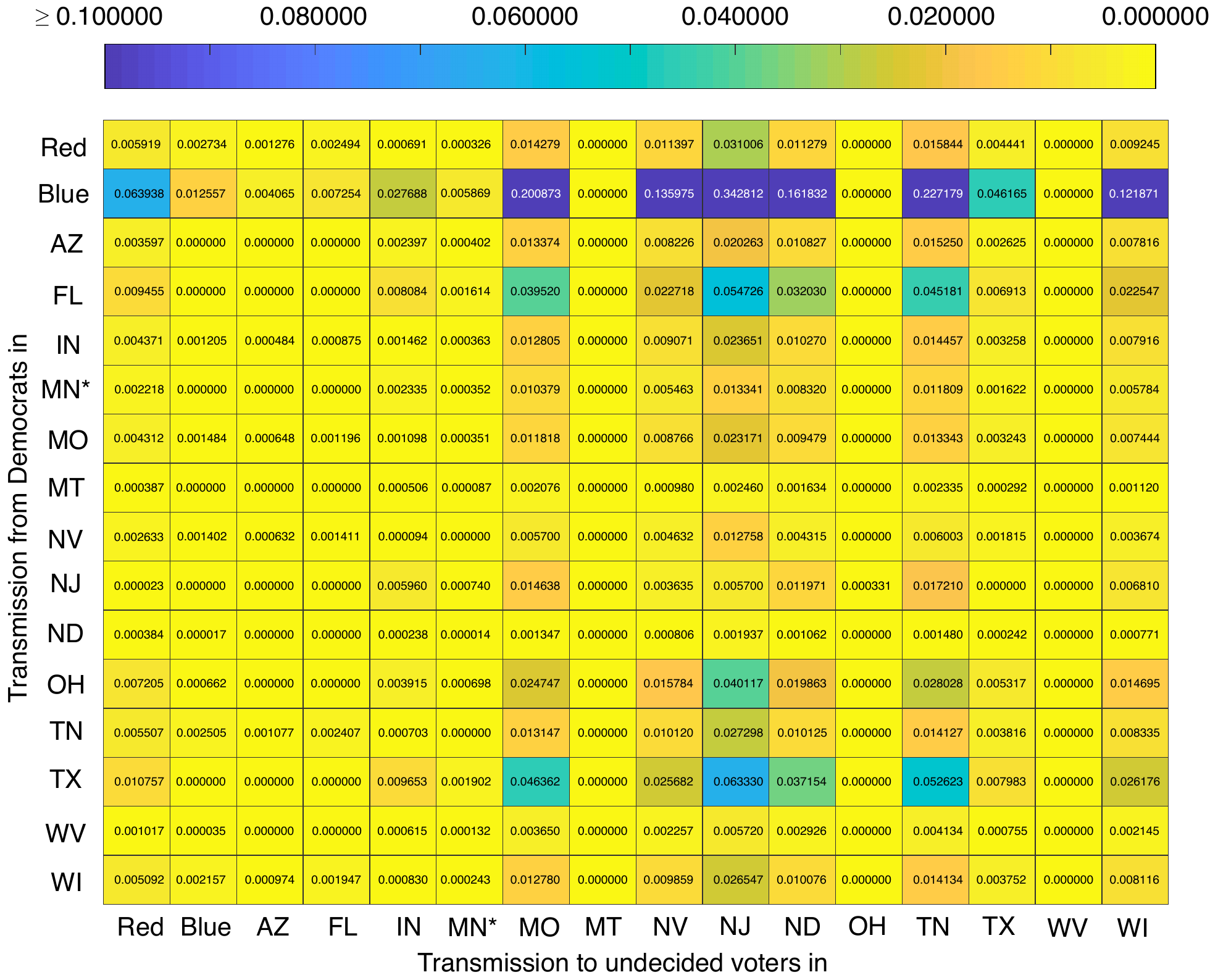}
\caption{{Democrat transmission parameters for our 2018 senatorial forecasts.} We give the parameters {that we use in our stochastic model~\eqref{eq:sde1}--\eqref{eq:sde3} to simulate} elections for our final forecasts in our GitLab repository  \cite{Gitlab_elections}. {(In this figure, we show the parameters to $6$ decimal places; see  \cite{Gitlab_elections} for more precise numbers.)} As an example, this figure shows the Democrat transmission 
parameters {(in units of 1/month)} for our 2018 senatorial forecasts using polling data that we obtained from \cite{RealClearPoliticsData} through 3~November 2018.  {We calculate these parameters by fitting our deterministic model~\eqref{eq:elec1}--\eqref{eq:elec3} to polling data. To simulate elections, we use these parameters in our stochastic model~\eqref{eq:sde1}--\eqref{eq:sde3}.} The parameter $\beta_\text{D}^{ij}$ describes opinion transmission from Democrats in state $j$ to undecided voters in state $i$. \label{fig:par2}}
\end{figure}

\begin{figure}[t]
\centering
\includegraphics[width=\textwidth]{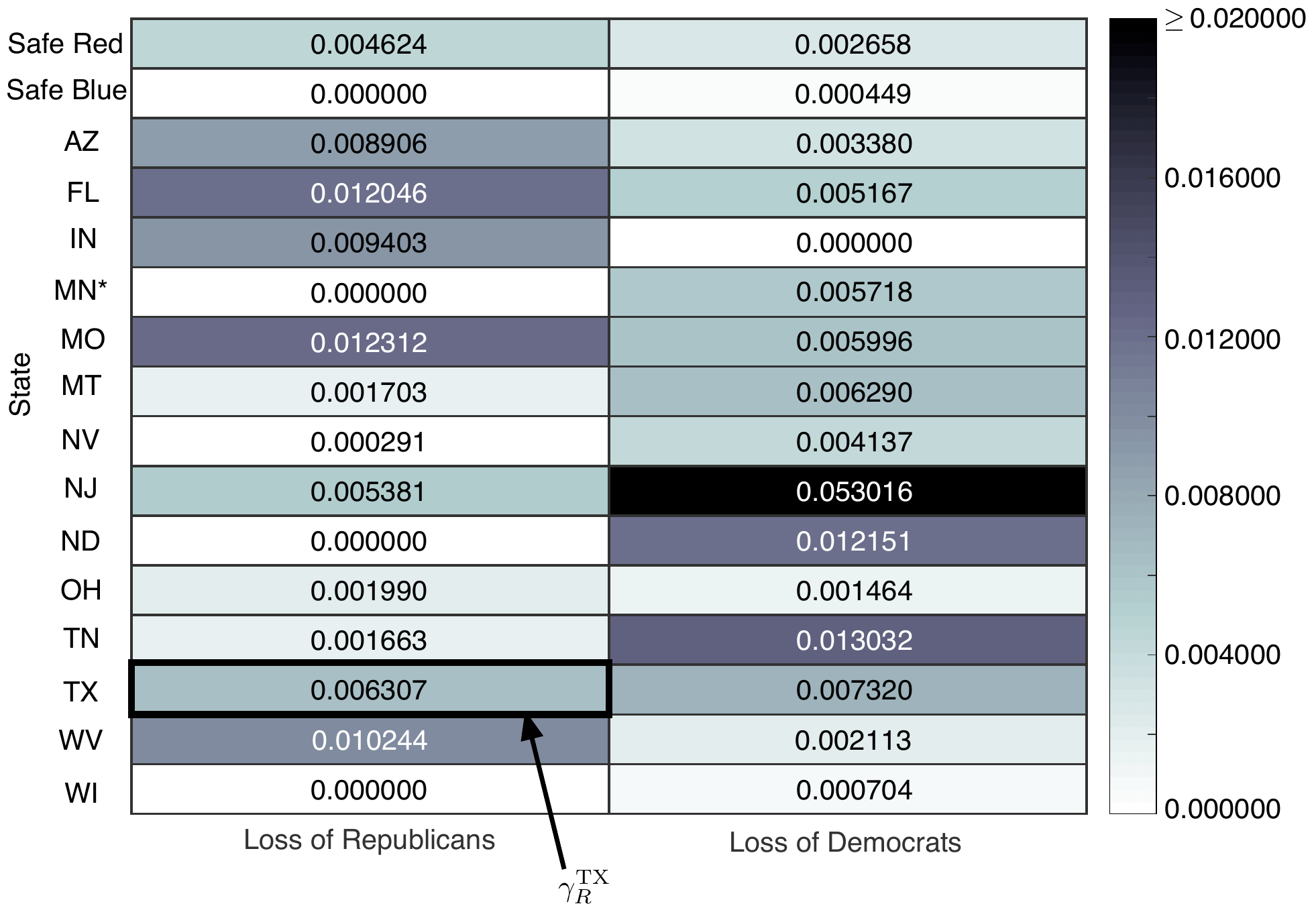}
\caption{{Turnover parameters for our 2018 senatorial forecasts.}
{As an example of our turnover parameters (e.g., the recovery parameters $\gamma^i_\text{R}$ and $\gamma^i_\text{D}$) in our models~\eqref{eq:elec1}--\eqref{eq:sde3}, we show the Republican and Democrat turnover parameters that we use in our 2018 simulations for the senatorial races. (We base these parameters on polling data that we obtained from \cite{RealClearPoliticsData} through 3~November 2018.)} The first column gives the $\gamma_\text{R}^i$ parameters, which describe the rate at which Republicans become undecided (in units of 1/months), and the second column gives the $\gamma_\text{D}^i$ parameters, which describe the rate at which Democrats become undecided (again in units of 1/months). To further clarify our notation, note that $\gamma_\text{R}^\text{TX}$ describes the rate at which Texas Republicans become undecided. These parameters suggest that committed voters do not typically change their minds during an election year. One can interpret $1/\gamma_x^i$ (where $x \in \{\text{D},\text{R}\}$) as the mean time that ``committed'' voters stay committed to their opinion before becoming undecided \cite{IdeaSpread}. Our parameter values suggest that the least committed voters in the 2018 senatorial races were New Jersey Democrats, yet $1/\gamma_\text{D}^\text{NJ} \approx 1.57$~years is still a long time on the scale of one election. {We show the parameters to $6$ decimal points; for additional precision (and the parameters that we use to simulate other elections), see  \cite{Gitlab_elections}.} 
\label{fig:par3} }
\end{figure}

\section*{Acknowledgements}

We thank Heather Zinn Brooks, Catherine Calder, Samuel Chian, Eli Fenichel, Serge Galam, William He, Brian Hsu, Kyle Kondik, Christopher Lee, Niall Mangan, Subhadeep Paul, Laura Pomeroy, and Samuel Scarpino for helpful comments and pointers to references. A.V. thanks Maciej Pietrzak for his advice on APIs. The work of A.V., D.F.L., and G.A.R. was supported in part by the Mathematical Biosciences Institute and the National Science Foundation (NSF) under grant no. DMS-1440386.  G.A.R. was also supported by  the NSF  grant no. DMS-1853587. A.V.\ is currently supported by the NSF under grant no.\ DMS-1764421 and by the Simons Foundation/SFARI under grant no.\ 597491-RWC.} 

%%%%

%\bibliographystyle{siam}
%\bibliography{scibib}

%%%%%

\end{document}